# Image Enhancement Techniques for Quantitative Investigations of Morphological Features in Cometary Comae: A Comparative Study

Nalin H. Samarasinha[1] and Stephen M. Larson[2]

[1] Planetary Science Institute, 1700 E Ft Lowell Road, Suite 106, Tucson, AZ 85719.

[2] Lunar and Planetary Laboratory, University of Arizona, 1629 E University Blvd, Tucson, AZ 85721.

47 Pages
15 Figures
0 Table

Corresponding Author: Nalin H. Samarasinha (nalin@psi.edu)

# Abstract

Many cometary coma features are only a few percent above the ambient coma (i.e., the background) and therefore coma enhancement techniques are needed to discern the morphological structures present in cometary comae. A range of image enhancement techniques widely used by cometary scientists is discussed by categorizing them and carrying out a comparative analysis. The enhancement techniques and the corresponding characteristics are described in detail and the respective mathematical representations are provided. As the comparative analyses presented in this paper make use of simulated images with known coma features, the feature identifications as well as the artifacts caused by enhancement provide an objective and definitive assessment of the various techniques. Examples are provided which highlight contrasting capabilities of different techniques to pick out qualitatively distinct features of widely different strengths and spatial scales. On account of this as well as serious image artifacts and spurious features associated with certain enhancement techniques, confirmation of the presence of coma features using qualitatively different techniques is strongly recommended.

# 1. Introduction

Characterization of many physical parameters of cometary nuclei (e.g., rotational states, locations of jet activity) and quantification of physical processes occurring in nuclei are based on spatial and temporal observations of anisotropic emission features in the cometary comae. However, in many instances, these coma features are only a few percent of the "coma background" and require enhancement of features to unambiguously identify them, to make measurements, and to carry out subsequent detailed analyses. Enhancement of images requires application of image enhancement techniques[1] where image enhancement could be defined as application of any type of operation(s) that improve the capability to detect features or to quantitatively characterize them in an image (e.g., Jähne 2004).

Image enhancement has many applications in a broad range of subjects encompassing physical and biological sciences, medical science, engineering, and computer science. Therefore the enhancement techniques developed in these fields are naturally designed for feature identifications and analyses optimized towards the specific types of features present. A wide array of enhancement techniques is described in the literature pertaining to image analysis in these diverse fields (e.g., Jähne 2004, O'Gorman et al. 2008, and references therein). However, many of these methods may be of little or no relevance to the analysis of coma morphologies of comets due to the nature of the brightness distribution and the type of features present in the cometary comae. In general, a coma has a radial fall-off of brightness from the optocenter (generally assumed to be the cometary nucleus) and many of the enhancement techniques that are appropriate for enhancing features in the coma are capable of minimizing the radial component and increasing the contrast of the features. The techniques discussed in this paper are relevant for coma observations when the nucleus is not resolved, especially ground-based images but works well for some space-based (e.g., Hubble Space Telescope) images too.

Historically, a range of enhancement techniques was used for the analysis of coma structures and is described in the literature (e.g., Larson and Slaughter 1992, Schleicher and Farnham 2004, and references therein). These references offer an excellent introduction to the basic capabilities of different image enhancement techniques used in the context of comets. In this paper, with numerically simulated images we characterize the pros and cons of enhancement techniques that are widely used in interpreting cometary coma observations and make a comparative analysis. We identify the techniques that are most suitable for the purpose of making measurements in the post-enhancement images. We also highlight the major drawbacks of some widely used enhancement techniques that may result in incorrect characterizations of the coma or nuclear phenomena. We do not intend to review all the techniques that have been used in the literature, but those best suited for quantitative studies among the widely used.

---

[1] In the literature, some authors use the phrases "image enhancement" and "image processing" interchangeably. However, we prefer to use "image enhancement" to mean operations that improve the capability to detect features whereas "image processing" is reserved to describe other operations on images such as standard image reduction tasks (e.g., bias subtraction and flat fielding), co-adding of images, cosmic-ray removal etc.

Section 2 of this paper characterizes various enhancement techniques and includes detailed descriptions for the techniques. In Section 3, comparative objective analyses between different techniques are made using numerically simulated images with known features. It also includes an assessment of the techniques for making measurements in post-enhancement images. Section 4 highlights the summary and conclusions of this paper. Appendix A derives the spatial distribution of flux for idealized cases and Appendix B provides an error analysis for selected enhancement techniques.

## 2. Characterizations of Widely Used Coma Enhancement Techniques

The structures in the coma are indicative of anisotropic emission of gas and dust from the nucleus. Therefore, correct identifications and measurements corresponding to spatial information of coma structures are needed for accurate interpretations of the coma observations. In this context, comparisons between different enhancement techniques and what changes they introduce to the original unenhanced images are relevant. In other words, one needs to ascertain the enhancement techniques that preserve the spatial information of coma features for accurate assessment of feature characteristics. This is critical when one makes measurements using post-enhancement dust or gas images to determine parameters such as the respective outflow velocities of dust or gas in the coma. Naturally, making measurements using the unenhanced original images is the ideal scenario. However, for many instances, unenhanced images themselves may not have sufficient contrast to identify specific distinguishing markers on the coma features thus necessitating one to use enhanced images to make the measurements.

Some enhancing techniques may emphasize certain types of features and not detect other kinds of features at all (also see Schleicher and Farnham 2004). In other words, the capability to unambiguously identify a particular kind of a coma feature could be specific to a select group of enhancement methods. Also, a given enhancement technique may introduce a set of enhancement artifacts specific to that technique only. Considering all these, it is critically important to understand the effects of a given technique on the resultant enhanced image and one should be cautious of over-interpretations. In addition, the brightness level (in comparison to the background) of the features identified using an enhancement technique should be comparable or better than the corresponding signal-to-noise in the original image. For example, if the signal-to-noise is 100, the feature should be about 1% or more above the background. Any feature fainter than this is likely to be an image artifact.

In the context of feature identification and measurements in the cometary coma, the majority of the widely used enhancement techniques fall into several categories. They are: (a) simple contrast stretches that show the presence of coma features, (b) the techniques that show features by explicitly removing a gross background, (c) the methods that enhance spatial discontinuities via spatial filtering, and (d) the techniques that enhance spatial discontinuities by means of spatial derivates. It should be pointed out that many individual techniques that search for spatial discontinuities might also remove a gross background. In many cases, this can create unintended and misleading results such as removing a "broad" coma feature but highlighting the low-contrast substructure therein as shown in Section 3 (ref. Figure 8) — a textbook example of missing the forest for the trees! It should be noted essentially almost all enhancement techniques do not preserve the flux distribution of the images. In addition, depending on the technique, it may enhance certain types of features while not detecting other features at all. Therefore, one should be cautious of over-interpretations.

Different enhancement techniques listed under a given category may produce different post-enhancement products. However, in general, as it will be shown later in Section 3, the images enhanced using these different techniques that fall into a given category produce certain common characteristics.

## 2.1 Simple Techniques that Find the Presence of Coma Features

These techniques typically involve a single straightforward operation on the image(s) and generally do not require complex algorithms and software.

**(a) Non-linear contrast stretches**

Perhaps the most widely used yet simple, non-linear contrast stretch is displaying the image using a logarithmic scale. This will allow one to display a high dynamic range image and identify the presence of coma features. As this does not alter the flux distribution of an image but only displays it in a logarithmic scale, there is no danger of introducing spurious features or artifacts into the image. Such contrast stretches are extremely useful for validating the features identified with more complex techniques described later in Section 2. However, in some cases, the contrast stretches might not be sufficiently sensitive to detect low-contrast features and they might be identifiable only with the more complex techniques described later in this paper.

Mathematically, the logarithmic display satisfies the condition
$$I_{out}(x,y) = \log(I_{in}(x,y)) \tag{1}$$
where the $(x,y)$ rectangular coordinate system has the nucleus/optocenter as its origin and $I_{out}(x,y)$ and $I_{in}(x,y)$ represent respective output and input pixel values. Throughout this paper, the location of the nucleus and the symbols for input and output images are represented as described above unless specifically stated otherwise.

Additional non-linear stretching techniques include square root, square, and sinh functions as well as histogram equalization. The histogram equalization effectively spreads out most frequent brightness values since pixel brightnesses will be equalized among all the histogram bins.

**(b) Temporal derivative**

This technique simply involves the difference (either by division or subtraction) of two images of the same object taken at different times (typically this time interval is of the order of a few minutes to a few hours and corresponds to a timescale over which spatial movement of coma features is of the order of or greater than the seeing disk of the observations). If the features have moved by a recognizable distance during the corresponding time interval, the resultant image will have corresponding positive and negative signatures. The positive-negative pairs are a telltale sign of the presence of real features.

If the early image is subtracted from the late image, the resultant image pixels are represented by

$$I_{out}(x,y) = I_{late}(x,y) - I_{early}(x,y) \tag{2}$$

whereas if the late image is divided by the early image, then

$$I_{out}(x,y) = \frac{I_{late}(x,y)}{I_{early}(x,y)} \tag{3}$$

where $I_{out}(x,y)$ represents the output pixel values while $I_{late}(x,y)$ and $I_{early}(x,y)$ represent respective pixel values corresponding to late and early images. Before image subtraction or division, it is necessary to register images so that the optocenters and the image orientations should be the same. Typically, the positive signatures in a positive-negative pair in the enhanced image would be further away from the nucleus than the corresponding negative signatures as most features move away from the nucleus with time. Temporal derivatives, for example, are suitable for detecting features that move away from the nucleus with time (e.g., CN shells; Schulz 1992, fast moving ion features). Figure 1 illustrates the usage of temporal derivative to demonstrate the motion of coma features.

The enhanced image due to the temporal derivative has higher relative noise (i.e., noise-to-signal ratio) than either of the original images. If the noise levels in each of the original images are comparable, the division will result in a relative noise approximately $\sqrt{2}$ times higher than either of the original images. In the case of the subtraction, the increase in the relative noise is even larger.

## 2.2 Techniques that Find Coma Features by Explicitly Removing a "Background"

In most cases, either subtracting or dividing the original image by an image representative of the background accomplishes the explicit removal of an approximation of the background. In general, for the techniques described in this Section, division by a background is preferred over subtraction due to practical considerations such as uncertainties associated with properly scaling the pixel values. Also, when far from the nucleus and the coma brightness is relatively smaller, subtraction of a small background value from the small coma brightness value results in another small value. Therefore, definite identification of features far from the nucleus becomes harder. These and other arguments listed later in this Section lead us to favor division over subtraction. In the discussion to follow, the emphasis is placed on division by the background and only the corresponding mathematical expressions are listed.

All the techniques discussed in this section that involve an image division require removal of sky contribution before application of the enhancement technique (especially if the sky level is significant when compared with the total flux in the regions where one prefers to search for coma features). For the techniques that involve subtracting the background after calculating it directly from the image itself, lack of an accurate sky removal will not be an issue. In that case, the subtraction will subtract out any residual sky contribution. On the other hand, one must carry out a careful sky subtraction if the technique will involve division by the background. If not, the resultant enhanced image will not be reliable especially at locations where the coma brightness is relatively smaller.

### (a) Division by or subtraction of an $1/\rho$ profile

The division or subtraction by a radial profile with an $1/\rho$ fall-off where $\rho$ is the skyplane-projected cometocentric distance relies on the fact that for a spherically symmetric steady state outflow with constant velocity, the column density distribution (and hence the flux) varies as $1/\rho$ (Appendix A). Note that this profile is azimuthally symmetric. Near-nucleus dust comae of many comets can be approximated by such a spatial profile and therefore provide a convenient yet reasonable model to divide or subtract out the gross background. On the other hand, asymmetric dust emission towards the sun will leave bright and faint areas, while images of gas comae show much flatter spatial profiles near the nucleus. Therefore division or subtraction by a $1/\rho$ profile may not be the most appropriate technique for enhancing the faint gas features. For this reason, for a gas coma, division or subtraction by an $1/\rho$ profile may yield an artificial "hole" around the nucleus region indicating that an excessive background was assumed there.

The division by an $1/\rho$ profile can be expressed as
$$I_{out}(x,y) = I_{in}(x,y)\rho(x,y) \tag{4}$$
with $\rho^2 = x^2 + y^2$ \hfill (5)
where the $(x,y)$ rectangular coordinate system has the nucleus as its origin and $I_{out}(x,y)$ and $I_{in}(x,y)$ represent respective output and input pixel values.

In addition to the reasons described at the beginning of this Section, subtraction of the background results in larger relative noise (i.e., noise-to-signal ratio) when compared with the division by the background. The corresponding derivations of noise characteristics for the division by an $1/\rho$ profile and subtraction of an $1/\rho$ profile are given in Appendix B.

### (b) Division by or subtraction of azimuthal average

In this technique, division by or subtraction of an azimuthally averaged radial profile is carried out. This profile is calculated by azimuthally averaging the pixels at a given $\rho$ for all $\rho$. The $\rho$ dependence of an image generated by an azimuthally averaged profile could be close to but not exactly the same as an $1/\rho$ image. Especially in the case of gas species and for non-spherically symmetric outflows, the azimuthally averaged profile will generate a more representative "background" than an $1/\rho$ image.

The division by azimuthal average can be expressed by

$$I_{out}(x,y) = \frac{I_{in}(x,y)}{I_{aziave}(\rho)} \qquad (6)$$

with $I_{aziave}(\rho) = \dfrac{\sum_{i,j} I_{in}(x_i, y_j)}{n_\rho}$ (7)

and $\rho^2 = x^2 + y^2 = x_i^2 + y_j^2$ . (8)

The quantity $n_\rho$ is the number of pixels satisfying equation (8) for a given $\rho$.

An inappropriate value for the azimuthal average for a given $\rho$, for example introduced due to the presence of a bright star, can be avoided by adopting a suitable clipping algorithm to exclude pixel values beyond a certain deviation from the mean pixel value.

**(c) Division by or subtraction of azimuthal median**

Division by or subtraction of azimuthal median is qualitatively similar to the division by or subtraction of azimuthal average. When there are a large number of anomalous pixels present (e.g., cosmic ray hits, hot pixels, or dead pixels), azimuthal median will provide a more representative "background" than the azimuthal average. This "background" is azimuthally symmetric.

The mathematical representation for division by azimuthal median can be expressed by

$$I_{out}(x,y) = \frac{I_{in}(x,y)}{I_{azimed}(\rho)} \qquad (9)$$

with $I_{azimed}(\rho) = median\,(I_{in}(x_i, y_j))$ (10)

where $\rho$, $x$, $y$, $x_i$, and $y_j$ are related by equation (8).

**(d) Division by or subtraction of azimuthal minimum**

Division by or subtraction of azimuthal minimum also removes a "background" which could be considered the "lowest common background" and produces enhanced images that are close to (b) and (c) above. However, care should be taken so that the minimum azimuthal value is not affected by false-minima such as dead pixels, for example by adopting a clipping algorithm. Subtraction of azimuthal minimum has been adopted by Lederer et al. (2009) to analyze images of comet Hale-Bopp (C/1995 O1). The "background" calculated here is again azimuthally symmetric.

Mathematically, one may represent the division by azimuthal minimum by

$$I_{out}(x,y) = \frac{I_{in}(x,y)}{I_{azi\min}(\rho)} \tag{11}$$

with $I_{azi\min}(\rho) = \min(I_{in}(x_i, y_j))$ (12)

where $\rho, x, y, x_i,$ and $y_j$ are related by equation (8).

**(e) Azimuthal renormalization**

In azimuthal renormalization[2] (Klinglesmith 1986), pixels at a given radial distance $\rho$ from the nucleus (optocenter) are azimuthally renormalized. The renormalization is carried out in such a way to guarantee that post enhancement, for all $\rho$, the respective minimum and maximum pixel values are $i_{min}$ and $i_{max}$. Generally, $i_{min}$ is chosen to be zero.

The azimuthal renormalization is expressed by

$$I_{out}(x,y) = i_{\min} + (i_{\max} - i_{\min}) \frac{I_{in}(x,y) - I_{\min}(x_i, y_j, \rho)}{I_{\max}(x_i, y_j, \rho) - I_{\min}(x_i, y_j, \rho)} \tag{13}$$

where $I_{\min}(x_i, y_j, \rho)$ and $I_{\max}(x_i, y_j, \rho)$ are the minimum and maximum pixel values for a given $\rho$ in the original image and $\rho, x, y, x_i,$ and $y_j$ are related by equation (8). Again, adopting a clipping algorithm to exclude outlier pixel values due to bright stars, cosmic rays, dead pixels, etc. is strongly advocated.

**(f) Division by or subtraction of an average or median "mask" based on images over a suitable time interval (e.g., covering a full rotation period)**

When the coma morphology varies in a periodic fashion, for example as a response to rotation of the nucleus, one may create an average or median "mask" using a series of images to represent the "background" (e.g., Schleicher and Woodney 2003). Then each image in the sequence can be either divided or subtracted by this mask. For this technique to be successful (i.e., for the mask to be representative of the background), the observing geometry should be nearly constant over a rotation cycle and the observational sampling of the images should be such that the images are approximately equi-spaced in time, sufficiently dense in temporal coverage (preferably ~10 images or more), and cover approximately a rotational cycle. Unlike the other techniques described in Section 2.2, where the calculated background is azimuthally symmetric, the background in this technique is not azimuthally symmetric but could be thought of as more representative of the actual background. Furthermore, spatially invariant coma features during a rotation cycle (e.g., the dust tail, time invariant polar jets, etc.) will be counted as part of the "background" and therefore they will be missing in the final enhanced image. This technique further requires some *a priori* knowledge about the comet's rotation period, whereas other techniques do not. Therefore, it cannot be safely applied before some preliminary studies have been conducted on the rotation period of the comet.

---

[2] Azimuthal renormalization and variations of it are referred to as "ring masking" by some authors in the early literature (e.g., Schwarz et al. 1989).

Mathematically, the division by an average mask can be expressed by

$$I_{out,k}(x,y) = \frac{I_{in,k}(x,y)}{I_{Nave}(x,y)} \tag{14}$$

with $I_{Nave}(x,y) = \dfrac{\sum_{k=1}^{N} I_{in,k}(x,y)}{N}$ \hfill (15)

where $I_{out,k}(x,y)$ and $I_{in,k}(x,y)$ represent the output and input pixel values respectively for the $k^{th}$ image. The number of images in the series, which defines the masked image, is $N$ and all the images in the series should be registered such that their origins are at the respective optocenters and have the same orientation. Adopting a clipping algorithm could be useful for determining the average mask.

Similarly, division by a median mask can be expressed by

$$I_{out,k}(x,y) = \frac{I_{in,k}(x,y)}{I_{Nmed}(x,y)} \tag{16}$$

with $I_{Nmed}(x,y) = median\ (I_{in,k}(x,y))$ \hfill (17)

where $k=1, N$.

## (g) Division by or subtraction of an azimuthally averaged or an azimuthally median filtered "mask" based on images over a suitable time interval (i.e., covering a full rotation period)

As a variation of the previous technique, one can redefine the mask by considering the azimuthal average or the azimuthal median of the series of images. Unlike the previous enhancement technique, here one should be able to recover spatially invariant features too as the background will be azimuthally symmetric.

The mathematical representation for the division by an azimuthally averaged mask is expressed by

$$I_{out,k}(x,y) = \frac{I_{in,k}(x,y)}{I_{Naziave}(\rho)} \tag{18}$$

with $I_{Naziave}(\rho) = \dfrac{\sum_{i,j} I_{Nave}(x_i, y_j)}{n_\rho}$ \hfill (19)

where $\rho, x, y, x_i$, and $y_j$ are related by equation (8) and $n_\rho$ is defined just after equation (8). $I_{Nave}$ has the same meaning as in equation (15). Adopting a clipping algorithm could be useful for determining the average mask.

For comparison, division by an azimuthally median filtered "mask" is expressed by

$$I_{out,k}(x,y) = \frac{I_{in,k}(x,y)}{I_{Nazimed}(\rho)} \tag{20}$$

with $I_{Nazimed}(\rho) = median\ (I_{Nmed}(x_i, y_j))$ \hfill (21)

where, $\rho$, $x$, $y$, $x_i$, and $y_j$ are related by equation (8) and $I_{Nmed}$ is similar to that defined in equation (17).

The techniques that subtract (or divide) an azimuthally constant background preserve the relative brightnesses of features at a given skyplane-projected cometocentric distance $\rho$. However, note that the radial profile of this background could easily deviate from the actual "background" and therefore one must be extremely cautious in interpreting radial behavior of features such as an apparent decrease in brightness of a jet at a certain radial distance from the nucleus. All enhancement techniques described in this section except for the techniques described under (a), (f), and (g), are not capable of detecting any azimuthally symmetric features such as shells since such features are considered part of the "background". Therefore, in order to search for such azimuthally symmetric features, it is instructive to inspect the azimuthally symmetric "background" corresponding to techniques (b), (c), and (d). Presence of shells for example, would show up as relative inflections or slow-down of the fall-off of "background" flux as a function of $\rho$. Note that technique (e) does not explicitly evaluate a "background" although it does "remove" an effective "background".

## 2.3 Techniques that Identify Coma Features by Exploring Spatial Discontinuities Via Spatial Filtering

**(a) Constant spatial filtering techniques such as unsharp masking**

Constant spatial filtering techniques identify coma features by bringing out spatial discontinuities. Perhaps the most widely used spatial filtering technique is a variation of the classical photographer's unsharp masking. Here, convolving with a two-dimensional Gaussian function (e.g., using the *gauss* task in Image Reduction and Analysis Facility (IRAF) of the National Optical Astronomy Observatory (NOAO)) will smooth the original image and then the original image will be divided (or subtracted) by the smoothed image.

This can be expressed by
$$I_{out}(x,y) = \frac{I_{in}(x,y)}{I_{in}(x,y) * g(x,y)} \tag{22}$$
and $I_{in}(x,y) * g(x,y) = \int_{-\infty}^{\infty} \int_{-\infty}^{\infty} I_{in}(x-\tau_x, y-\tau_y) g(\tau_x, \tau_y) d\tau_x d\tau_y$ (23)

where $g(x,y)$ is the two-dimensional Gaussian function and $I_{in}(x,y) * g(x,y)$ represents the convolution of the original image by the Gaussian function. The integral over the parameter $\tau_x$ indicates that the function $I_{in}$ is convolved with the Gaussian function $g$ in the direction $x$. I.e., for a given value of $x$, the integral evaluates the amount of overlap of $g$ with $I_{in}$ as $g$ is gradually moved across $I_{in}$ in the $x$ direction. Similarly, the integral over $\tau_y$ represents the convolution of $g$ and $I_{in}$ in the $y$ direction. Although the integrations over $\tau_x$ and $\tau_y$ are from -∞ to +∞, from a practical point of view, it is sufficient to carry out the integration from $-l$ to $+l$ where $l$ represents a few standard deviations of the Gaussian function, *g*. Typically, for *g*, a circular Gaussian function is chosen. Despite the fact that convolving with a two-dimensional Gaussian function is the most commonly used, one-dimensional convolving functions are referenced in the literature (e.g., Schwarz et al. 1989).

The spatial scales of features detected by this technique will depend on the FWHM of the smoothing Gaussian function selected. In other words, the features identified in the enhanced images are sensitive to the FWHM of the Gaussian. Therefore, care should be exercised to make sure that no features are missed due to this fact. Furthermore, the resultant image can be thought of as a negative of the second derivative in the spatial domain. For example, a point source such as a star will show up as a bright point surrounded by a negative ring whereas a linear spatial discontinuity will show up as positive-negative linear counterparts.

Related to unsharp masking is the more complex wavelet filter, which is available in many data processing packages. Here the enhanced image is constituted from a series of wavelet images created based on the original image. This technique has been applied for identifying features in cometary comae (e.g., Murtagh et al. 1993) but not as widely used as many other techniques described in this paper and suffers from some of the shortcomings inherent to the unsharp masking.

**(b) Radially variable spatial filtering**

This algorithm compensates for the fact that anisotropic features in the coma expand with time and distance from the nucleus due to particle and velocity dispersion. This implementation is a variable, abbreviated, kernel deconvolution, which is used to suppress intensity gradients larger than the kernel size (Larson and Slaughter, 1991; Schleicher and Farnham, 2004; McCarthy et al., 2007). By increasing the kernel size as a function of distance $\rho$ from a specified location $(x_0, y_0)$, such as the comet nucleus, it is possible to optimize visibility of spatial features that become intrinsically larger due to velocity dispersion (e.g., dust jets originating from the nucleus) as they move away from their source.

A logarithmic pixel value, $I(x,y)$, of the target pixel is altered by the Gaussian-weighted sums of a usually non-contiguous 3×3 orthogonal pixel matrix defined by the original (central) target pixel, 4 corner pixels, and 4 edge pixels (Figure 2). The separation of the edge pixels from the kernel center is given by $a$, and the corner pixels are $\sqrt{2}\,a$ away from the center. The weighting factors are consistent with a Gaussian profile and the resultant pixel value, $I_{out}(x,y)$, is given by

$$I_{out}(x,y) = 1024\, I(x,y) - 192\, (\Sigma\, I_{ep}) - 64\, (\Sigma\, I_{cp}) \qquad (24)$$

where $\Sigma$ denotes the summation over relevant pixel values and
$I(x,y)$ = original pixel value in log DN (digital number) of the target pixel at $(x,y)$,
$I_{ep}$ = edge pixel values in log DNs, and
$I_{cp}$ = corner pixel values in log DNs.

The coefficients 1024, 192, and 64 are representative of a Gaussian profile.

Separation of the edge pixels from the target pixel at location ($x,y$) is further given by

$$a = a_0 + b_0\, \rho^n \qquad (25)$$

where $\rho$ = skyplane-projected offset of ($x,y$) from the specified reference location ($x_0,y_0$),
$a_0$ = minimum kernel size at ($x_0,y_0$) (i.e., separation of an edge pixel from ($x_0,y_0$) when the offset $\rho$ =0),
$b_0$ = azimuth-independent slope of kernel size with $\rho$, and
$n$ = exponent allowing non-linear kernel size growth with distance $\rho$.

Typically, $a_0$ is of the order of a unit pixel width. As the parameter $a$ is azimuthally symmetric, it does not distinguish between sunward and anti-sunward directions. Note that the definition of the parameter $a$ is slightly different in Larson and Slaughter (1992) as the kernel size growth is linear with distance $\rho$ there.

To reduce artifacts caused by integer pixel shifts in the size of the kernel, bilinear interpolation between nearest pixels is used to determine the effective DNs of the edge and corner "pixel" locations.

As with many image enhancement products, the resulting image must be viewed with caution. It is a map of the rate of change in brightness over ever increasing spatial scales from the nucleus. Photometric information is not retained. Since there is no directional dependency in the visibility of brightness changes, both radial and circular features are represented. The best combination of parameters is usually derived empirically to optimally reveal the features. The enhanced images serve as a guide to the location of features that can also be seen (often with some difficulty) in unenhanced images with appropriate contrast stretch. This enhancement allows features over a wide range of brightness to be seen at the same time.

## 2.4 Techniques that Identify Coma Features by Exploring Spatial Discontinuities with Spatial Derivatives

### 2.4.1 First Derivative Techniques

The following techniques identify coma features via spatial discontinuities existing in images; these techniques are seeking and highlighting the spatial discontinuities by applying linear or rotational shifts to images.

**(a) Linear shift differencing (either in one dimension or in two dimensions)**

The most basic form of the linear shift differencing involves shifting the image linearly in one dimension and then subtracting the shifted image from the original. Mathematically this uni-directional shift differencing can be expressed by

$$I_{out}(x,y) = I_{in}(x,y) - I_{in}(x+\Delta x, y) \tag{26}$$

where $\Delta x$ is the shift applied in the *x*-direction (the shift here was arbitrarily chosen to be in the *x*-direction but could be in any direction). The spatial discontinuities that are comparable in magnitude to $\Delta x$ will show up as positive-negative pairs. These discontinuities are preferentially parallel to the shift. A better alternative is to apply shifts in linearly opposite directions (bi-directional linear shift differencing) so there will be no positive-negative pairs and the locations of features are easier to determine. The relevant mathematical expression is

$$I_{out}(x,y) = [I_{in}(x,y) - I_{in}(x+\Delta x, y)] + [I_{in}(x,y) - I_{in}(x-\Delta x, y)]. \tag{27}$$

The corresponding expression when the shifts are applied in the orthogonal but linearly opposite directions is given by

$$I_{out}(x,y) = [I_{in}(x,y) - I_{in}(x, y+\Delta y)] + [I_{in}(x,y) - I_{in}(x, y-\Delta y)]. \tag{28}$$

Application of bi-directional shifts both in *x* and *y* (i.e., both operations given by equations (27) and (28)) will result in the two-dimensional application of the technique. This will cause less distortion to the locations of the features. This two-dimensional bi-directional linear shift differencing shares certain similarities with the unsharp masking technique described in Section 2.3.

An early application of the one-dimensional uni-directional (vertical, horizontal, or diagonal) shift differencing of cometary images of comet 1P/Halley is discussed in Klinglesmith (1981).

## (b) Rotational shift differencing

The uni-directional rotational shift differencing (i.e., rotation of the original image, typically around the optocenter, which presumably represents the nucleus, and then subtracting that image from the original image) will indicate spatial discontinuities that are of the order of the rotation angle $\Delta\theta$ as positive-negative pairs. Therefore, accurate identifications of feature locations in the resultant image may not be straightforward. The spatial discontinuities identified are preferentially in the azimuthal direction. Mathematically, this operation can be expressed by

$$I_{out}(\rho,\theta) = I_{in}(\rho,\theta) - I_{in}(\rho,\theta+\Delta\theta) \tag{29}$$

where $\theta$ is the azimuth angle (measured counterclockwise from the +*x* axis) and is related to (*x,y*) by the equations

$$x = \rho\cos\theta \tag{30}$$

and

$$y = \rho\sin\theta. \tag{31}$$

A more appropriate approach would be to use bi-directional rotational shift differencing given by

$$I_{out}(\rho,\theta) = [I_{in}(\rho,\theta) - I_{in}(\rho,\theta+\Delta\theta)] + [I_{in}(\rho,\theta) - I_{in}(\rho,\theta-\Delta\theta)]. \tag{32}$$

### (c) Radial shift differencing (either a constant shift or a shift proportional to $f(\rho)$)

The radial shift differencing, could be useful for identifying radial discontinuities. The radial shift differencing with a constant shift can be mathematically expressed by

$$I_{out}(\rho,\theta) = I_{in}(\rho,\theta) - I_{in}(\rho + \Delta\rho, \theta). \tag{33}$$

If the radial shift is proportional to a function of $\rho$, $f(\rho)$, we have

$$I_{out}(\rho,\theta) = I_{in}(\rho,\theta) - I_{in}(\rho + kf(\rho), \theta) \tag{34}$$

where $k$ is a constant. The rationale for using such a functional dependence for the shift for $\rho$ is that the radial dispersion of features in the coma increases with $\rho$.

### (d) Simultaneous rotational and radial shift differencing

By incorporating the operations in equations (32) and (34), a simultaneous rotational and radial shift differencing can be defined and the relevant mathematical expression is

$$I_{out}(\rho,\theta) = (I_{in}(\rho,\theta) - I_{in}(\rho + kf(\rho), \theta + \Delta\theta)) + (I_{in}(\rho,\theta) - I_{in}(\rho + kf(\rho), \theta - \Delta\theta)). \tag{35}$$

Note that this expression shares many similarities, but is not equal, to the equation (1) given in Larson and Sekanina (1984) and a more detailed description, including enhanced images for slight variations of this technique can be found there. Also, this technique shares certain similarities with the radially variable spatial filtering discussed in Section 2.3(b). However, they are not identical.

For these techniques discussed in Section 2.4.1, the magnitude of the shifts used in a particular technique determines the relevant spatial scales of the features that can be detected. For example, the discontinuities with spatial scales much larger than or much smaller then the shift applied could easily escape detection. For the techniques described above, the shift(s) could be applied either in one direction or in opposite directions. When the shift is applied in one direction, the discontinuities, preferentially parallel to the shift, will show up as positive-negative pairs. Such a positive-negative pair suggests the existence of a spatial discontinuity; however, a detailed characterization of the feature is difficult. Applying the shift in opposite directions will generate an enhanced image with no such positive-negative pairs. By applying the shifts also in the orthogonal direction, one can search for discontinuities in two orthogonal directions (i.e., in two dimensions) and may provide a more accurate picture. Despite that, as mentioned earlier, the magnitude of the shift(s) may still preclude detection of all features present in an image.

All the techniques in this section 2.4.1 have increased noise in the enhanced image. For example, for the bi-directional linear shift differencing, the noise increases by about a factor of 2 and the relative noise increases by many factors when compared with the original unenhanced images. So, one should be careful in the interpretation of features.

### 2.4.2 Second Derivative Techniques

The Laplace filtering and its variations represent examples for second derivative techniques.

**Laplace filtering**

The Laplacian of an image is essentially the second derivative and the two-dimensional Laplacian can be mathematically expressed as

$$I_{out}(x,y) = \frac{\partial^2 I_{in}(x,y)}{\partial x^2} + \frac{\partial^2 I_{in}(x,y)}{\partial y^2}. \tag{36}$$

Generally, a Laplacian can be implemented using a convolution filter that approximates the second derivative. For example, in IRAF, the Laplace filter in the task *laplace* is represented by a set of four 3×3 kernels that approximate the Laplace operator. More evolved variants of this technique includes the adaptive Laplace filtering (e.g., Lorenz et al. 1993, Vincent 2010) available in the European Southern Observatory Munich Image Data Analysis System (ESO-MIDAS).

As the second derivative (and hence the Laplacian) will create negative values at the "peaks" in the image, one may consider the negative of the Laplace filtering as an appropriate alternative. However, as we will demonstrate later in Section 3, the Laplace filter can generate many image artifacts leading one to misidentify features. The Laplace filter is very sensitive to the noise and therefore the image should have extremely good signal-to-noise prior to application of this technique, as otherwise noise will get enhanced.

## 2.5 Other Techniques and/or Representations of the Images

There are additional enhancement products that may not strictly fall into any of the above categories; however, they may share certain characteristics or heritage from the above techniques. An example is the image showing a polar representation ($\rho,\theta$) (i.e., radius $\rho$ versus azimuth $\theta$) of an original image or an enhanced image.

Mathematically, a pixel ($x,y$) can be mapped to a pixel ($\rho,\theta$) of the polar image and $\rho$ can be expressed by equation (8) and $\theta$ is defined by equations (30) and (31) when $\theta$ is measured from the +x axis in a counterclockwise sense. When the origin of the $\theta$ axis is different, that corresponds to a different starting point for the azimuth and equations (30) and (31) must be altered appropriately.

The pixel values of the polar image are given by

$$I(\rho,\theta) = \sum_{1}^{n_{\rho,\theta}} \frac{I_{in}(x,y)}{n_{\rho,\theta}} \tag{37}$$

where $n_{\rho,\theta}$ is the total number of (x,y) pixels satisfying equations (8), (30), and (31) for given values of $\rho$ and $\theta$. Note that $\rho$=1 represents pixels surrounding the nucleus and corresponds to the very bottom row of the ($\rho,\theta$) image.

The polar display by itself is not strictly an enhancement technique, but it could facilitate feature characterizations and determination of locations of features. Sometimes, polar displays may result as a byproduct of specific enhancement techniques. The polar displays could also be helpful for identification of accurate position angles and the extent of the curvatures for jet features and apparent expansion speeds. Polar representations are especially useful in characterizing radial features (e.g., Schleicher et al. 2003, Knight et al. 2012, Mueller et al. 2013).

When converting an (x,y) image to a ($\rho,\theta$) image, a near-nucleus pixel in the (x,y) image will get mapped to many adjacent pixels in the $\theta$ axis. In other words, close to the nucleus, the brightness distribution along the $\theta$ axis may not be reliable. Therefore, especially close to the nucleus, binning the pixels in the radial direction can increase the reliability as well as the signal-to-noise of the brightness profile.

All the techniques in Section 2.2, which explicitly remove a background (but probably except technique (a) depending on how the algorithms are implemented) as an intermediate step of the enhancement process, may generate polar images. This is due to the fact that carrying out azimuthal operations is mathematically simpler on a ($\rho,\theta$) image than on a (x,y) image.

Another technique (or more appropriately that can be considered as an add-on to a different technique) is the application of an adaptive filter (e.g., Tom 1985). The idea is to appropriately adjust the filter used in another technique to achieve a certain objective. For example, the filter can be adjusted based on the locations of the corresponding pixels in the image or the brightness characteristics of the neighboring pixels. So, a filter can be varied such that the high contrast enhancements occur in regions with details while little or no enhancement occurs in smooth regions; the filter in this case could correspond to unsharp masking (e.g., Polesel et al. 2000). The Laplace filtering is another technique where an adaptive filter is used widely (see the discussion in Section 2.4.2; also Boehnhardt and Birkle 1994). Strictly speaking, one may argue even the radially variable spatial filtering technique described in Section 2.3(b) as an adaptive filter as the kernel of the filter increases as one moves away from the nucleus of the comet. An adaptive filter in most cases, if applied appropriately, will result in better outcomes than using the same underlying technique with a fixed filter. However, still it cannot eliminate all the shortcomings of the underlying technique.

# 3. Image Enhancement of Simulated Images

## 3.1 Graphical Representations

Graphical representations of enhancement techniques (or of different categories of enhancement techniques) are helpful for visualizations and thereby for explaining the mathematical operations carried out during the image enhancements. Furthermore, it aids the interpretation of the resulting enhanced images. Figure 3 is an example to demonstrate the relevant one-dimensional graphical representations for some of the techniques described in Section 2. It clearly demonstrates the characteristic signatures associated with the unsharp masking, first derivative techniques, and second derivative techniques. For example, the bottom panel of Figure 3 helps us visualize what occurs when a second derivative technique such as the Laplace filter is applied to an isolated single jet (Figure 4). This technique would create two maxima on either side of the minimum with the latter corresponding to the actual jet. Although one may argue that the inversion of the enhanced image in Figure 4 would simply yield the correct identifications of actual features, as we demonstrate later in Section 3.2, when multiple features are present in the coma, indeed there could be confusions as to what are actual features and what are artifacts. Unfortunately, such scenarios could lead to incorrect feature identifications and subsequent modeling of such mischaracterized (or misidentified) features.

## 3.2 Comparisons Between Different Techniques

As there are many enhancement techniques available, a natural question to ask is what technique (or techniques) should be used to identify coma features actually present in the original unenhanced images. Are there certain techniques that are preferred because they provide reliable information about the features present? Using simulated images, where we have *a priori* knowledge on the spatial distribution of coma features, accurate assessments can be made of the capabilities (as well as drawbacks) of different techniques to enhance and correctly identify low-contrast features.

In Figure 5(a), we show three simulated jet features created using the Monte Carlo coma simulation software developed by us. In Figure 5(b) this is co-added with a strong coma component generated by hemispherical emission (for example as a direct response to sunlight). As this hemispherical emission dominates that from the jets in panel (a), the three jet features are only barely visible in the highly stretched unenhanced image in panel 5(b). The rest of the panels show enhanced images for a range of enhancement techniques. It can be seen that except for the Laplace filter, all the techniques are capable of retrieving the three jets; however, it is clear that most techniques significantly affect the flux distribution in and around the jet features making "alterations" of varying degrees to the features. If one inverts the resultant image from the Laplace filter (i.e., if one considers the dark regions as the ones representing the jets), then the three jets can be identified.

Figure 6 again shows three jets and a strong asymmetric component due to hemispherical emission. However, now the rightmost jet is significantly weaker. It is evident based on the enhanced images that some techniques are not successful in detecting this weak jet feature and only the two brightest jets are recognizable. This demonstrates the difficulties encountered by some techniques to correctly identify actual features when the features are of widely different strengths — a scenario more common than what is shown in Figure 5. Such features of widely different strengths could be either intrinsic or due to particular observing geometries and for example such features were observed in comet 103P/Hartley 2 prior to the EPOXI encounter (e.g., weak CN feature close to the nucleus in row 4 of Figure 5 of Knight and Schleicher 2013 and the weak continuum feature in Figure 8 of Mueller et al. 2013).

To consider another complicated, yet realistic scenario, Figure 7 examines the case where two simulated intersecting (i.e., crisscrossing) jets are present in addition to a strong asymmetric "background" coma. It can be seen that some techniques fail to identify the intersecting nature of the jets.

Another scenario of interest is the presence of sub-structure in a relatively broad feature. Many near-nucleus coma images from space missions suggest that what is seen as individual features with minor sub-structure from the ground could actually be due to the relatively weak multiple jets or filaments emanating from the nucleus combined with the broader background/diffuse emissions (e.g., comet 1P/Halley; Keller et al. 1986, comet 19P/Borrelly; Soderblom et al. 2002, Boice et al. 2002, comet 9P/Tempel 1; A'Hearn et al. 2005, comet 81P/Wild 2; Sekanina et al. 2004, Farnham and Schleicher 2005, comet 103P/Hartley 2; A'Hearn et al. 2011). In other words, a single jet feature that can be easily identified in the ground-based images could be due to emissions from multiple yet relatively small source regions on the nucleus that are located in close proximity to each other or could be due to emissions from a relatively large source region with variable activity across that source region.

Figure 8 is a simulation corresponding to a strong broad jet with three weak narrow jets with the latter representing the sub-structure embedded in the broad jet. As can be seen from Figure 8, some enhancement techniques are adept at recovering the broad jet but not the narrow jets while the opposite is true of some other techniques. This highlights the complementary nature of these techniques and the danger of relying on a single technique. In many instances, especially for spatial filtering and spatial derivative techniques, the physical size of the features detectable is dependent on the shift or the kernel size adopted. Therefore, it is prudent to carry out the relevant image enhancement multiple times using different shifts or kernel sizes in order to be confidant that no features were missed.

In general, the techniques that "remove" a background coma discussed in section 2.2 are more adept at recovering the brighter broad jet whereas the techniques that are based on spatial discontinuities have a preference for detecting the weaker narrow jets. It is pertinent to point out that as the broad jet is the overwhelmingly bright feature in Figure 8 and therefore the dominant feature in this simulation (nearly two orders of magnitude brighter when compared with the weaker narrow jets), it is imperative that one should not fail to detect that. In that sense, the techniques that remove a background coma are preferred over techniques that are based on highlighting and detecting spatial discontinuities. However, to obtain the overall view and to reconstruct the actual scenario, one needs to use multiple enhancement techniques and the interpretation of the enhanced images and the identification of the features must be made with great care.

## 3.3 Image Artifacts Created During Image Enhancements

Another aspect of image enhancement that one must take great care to avoid is to make sure that artifacts created by an enhancement technique will not be misinterpreted as real features. Such artifacts in an enhanced image can be due to (a) image artifacts in the original image, which get amplified during the enhancement process, or (b) artifacts created during the enhancement process that are specific to that particular enhancement technique.

While many experienced observers could easily identify certain image artifacts in the original unenhanced images, it is necessary to bear in mind that, for example, even an incorrect flat-fielding or sky-subtraction which leaves a subtle gradient across an image could result in a bright broad jet-like feature in the enhanced image. Typically these types of "spurious features" are independent of the image enhancement technique applied (even though some techniques may highlight such artifacts more than the others). In most cases, there are telltale signs in the original image to indicate what may cause such image artifacts once enhanced.

On the other hand, artifacts or the effects caused by the enhancement technique are generally specific to the enhancement technique (or the category of techniques). For example, a classic outcome of the division by $1/\rho$ profile is the well-defined sharp "hole" at the center (at $\rho = 0$) surrounded by a small bright ring (e.g., panel (c) in Figures 5, 6, and 7). The hole at the center is caused by the fact that in the enhanced image, the central pixel is multiplied by $\rho = 0$. The diameter of the bright ring is of the same order as the FWHM of the effective point-spread-function. Note that the original simulated images were convolved with a Gaussian to mimic the astronomical seeing and therefore the flux at or near the center pixel is less than what corresponds to an $1/\rho$ profile.

For the techniques that require determination of the location of the nucleus, the coma structure near the nucleus in the enhanced images should be interpreted with caution. If the location for the nucleus selected is slightly offset from the actual location of the nucleus, this could result in false-features near the chosen location for the nucleus. To illustrate this point, in Figure 9, we enhance a simulated image with several different "nucleus locations" which are offset from the optocenter. It has been suggested that a slight offset between the optocenter and the actual location of the nucleus is possible. For example, Yeomans and Chodas (1989) based on the orbital fits to astrometric data, after taking non-gravitational effects into account, estimate that 1P/Halley's nucleus is offset by 880 km from the location of the optocenter when the comet is 1 AU from the sun. This offset for comet 1P/Halley is of the order of the astronomical seeing for many observations and Yeomans and Chodas have modeled the offset as proportional to the inverse square of the heliocentric distance. However, they do not quote offsets for other comets. More recent analyses also suggest that such systematic offsets have not been clearly detected in other comets (Chesley, S. 2013; personal communications). Such offsets could be caused by the presence of a strong coma feature in one hemisphere or observational artifacts generated by incorrect tracking or guiding.

In our own coma simulations, depending on the relative strengths of the features, we occasionally notice that the optocenter could be offset from the actual location of the nucleus by a small amount (typically of the order of a pixel or less). As seen from Figure 9, one should not rely on the morphology of features near the nucleus. The simulations shown in Figure 9 indicate that morphology within a region with an approximate radius an order of magnitude larger than the offset will be affected and the corresponding morphology therein is unreliable. Mueller et al. (2013) arrived at a similar conclusion for the effects on the radial spatial profiles near the nucleus when the location of the chosen nucleus is slightly different.

Based on our experience, we have noticed that sub-pixel sampling and corresponding weighting (e.g., in determining the appropriate $\rho$ for a given pixel) reduce image artifacts introduced during the enhancement, especially for the techniques that involve converting the rectangular coordinates into polar coordinates. To further elaborate, in this case, if a particular pixel $(x,y)$ is equally divided into $n \times n$ sub-pixels (for a total of $n^2$ sub-pixels) and only $m$ of those sub-pixels map into a certain $(\rho,\theta)$ pixel, the contribution from the pixel $(x,y)$ to the pixel $(\rho,\theta)$ is only a fraction $m/n^2$ of the pixel value of the pixel $(x,y)$. Therefore the effective sub-pixel weighted value of a pixel $(\rho,\theta)$ can be expressed as

$$I(\rho,\theta) = \sum_{i,j} f_{i,j} I(x_i, y_j) \tag{38}$$

where $f_{i,j}$ represents the fraction of sub-pixels of a pixel $(x_i, y_j)$ that maps into the pixel $(\rho,\theta)$ and one needs to sum over all relevant $(x_i, y_j)$ pixels.

Figure 10 shows the effects of sub-pixel sampling and weighting in the application of division by azimuthal average (top row) and azimuthal renormalization (bottom row). The images from left to right represent a simulated image, an enhanced image without sub-pixel sampling, and an enhanced image with sub-pixel sampling (with 10×10 sub-pixels for each pixel). It is clear that inadequate sampling causes the cross-like artifacts in the middle column of panels.

## 3.4 Making Measurements in Enhanced Images

In general, image enhancement techniques do not preserve the flux of each pixel in the original unenhanced image. So, direct flux measurements in enhanced images should be avoided. However, an exception to this is the azimuthal brightness profile of the enhanced image at a given skyplane-projected cometocentric distance, $\rho$, for the techniques (a), (b), (c), (d), (e), and (g) in Section 2.2. The azimuthal profiles for those techniques do not preserve the original fluxes, but the *relative flux for each pixel in ascending (or descending) order* is preserved thus enabling one to make qualitative comparisons between features at different azimuths (i.e., at different position angles). However, the order of relative fluxes of a radial profile of the corresponding enhanced images is not preserved.

On the other hand, images enhanced via many different enhancement techniques enable one to make measurements of spatial locations of coma features, which may not be possible with the original unenhanced images. It is relevant to know which enhancement techniques preserve the spatial locations of the features and therefore are able to provide accurate measurements thereof.

Simulated images provide an excellent opportunity to check which enhancement techniques would provide reliable estimates for the locations of features and therefore the determinations of dust and gas outflow velocities based on the accurate locations of coma features. Shown in Figure 11 are two scenarios of simulated Archimedean spirals corresponding to narrow and wide jets with a known model outflow velocity. Velocity determinations could be made based on the enhanced images. This was accomplished by measuring the radial outward motion of the features along diametrically opposite directions (e.g., up and down directions) in these and other enhanced simulated images (not shown) corresponding to later times. Both techniques, division by $1/\rho$ and application of a radially variable filter, resulted in extracted outflow velocities to within about 5% of the input value. The offsets between the input and extracted values were larger for the wide jet than for the narrow jet. These relatively larger offsets for the wide jet are caused by uncertainties associated with precise determinations of peak values in the features, as the flux distributions are broader for the wide jet.

## 3.5 Comparison with Techniques that Require Multiple Images

Next, we will explore whether techniques that require multiple images in order to perform the image enhancement, such as the technique (f) described in Section 2.2, could provide reliable results for the features. In the technique (f) of Section 2.2, the original image is divided by a masked image created by the average or the median of images covering a rotational cycle. The third row of Figure 12 shows a temporal sequence of enhanced images created by dividing the original images by a median mask generated from seven images taken over a rotational cycle. The bottom row of the same figure shows the corresponding images when the original images are divided by the respective azimuthal profiles. It can be seen that the two techniques provide consistent results.

The total number of images necessary to generate the median mask depends on the complexity of the coma structures present as well as on the relative strengths of the variable structures relative to a "nearly time-independent background coma." However, if the images are reasonably well sampled in rotational phase space and the relevant conditions are satisfied, as seen from Figure 12, even a small number of images are sufficient. Still, we caution that the features should be confirmed with another technique to make sure the morphological determinations are robust.

# 4. Summary and Conclusions

By enhancing coma images of comets 23P/Brorsen-Metcalf, 1P/Halley, and Hyakutake (C/1996 B2), Larson and Slaughter (1992) and Schleicher and Farnham (2004) demonstrate that not all enhancement techniques are equally capable of detecting different kinds of features present in the coma. In this paper, with simulated images where features that are actually present in the images are known *a priori*, we confirm that this is indeed the case. In addition, the simulated images provided us an opportunity to objectively assess the strengths and weaknesses of different coma enhancement techniques.

Therefore, there is no "one size fits all" enhancement technique capable of or suitable for detecting different coma features of varying spatial scales, morphology, and relative brightness. In general, the techniques which remove a background from the original image during the enhancement process are less likely to cause misleading image artifacts when compared with the techniques that search for spatial discontinuities, and especially second derivative techniques. In other words, the techniques that remove a common background are relatively benign for the purpose of detection of coma features as they are unlikely to cause serious image artifacts. However, they may not be effective in detecting weak features, which could be characterized as sub-structure of the strong features. The techniques which search for spatial discontinuities are more suited for detecting such weak features as those techniques look for spatial discontinuities and depend less on the relative strengths of the features. It is advisable to use more than one category of enhancement techniques and also, if possible, to confirm the presence of any such feature so identified with the help of multiple techniques by careful examining (or reexamining) of the original images with high contrast stretches.

Certain techniques, in particular the second derivative techniques such as the Laplace filter, could produce misleading post-enhancement images where identification of actual features from spurious features may not be obvious. Spatial filtering techniques as well as spatial derivative techniques are susceptible to missing actual features if the size of the filter or the kernel size chosen is not optimum for the actual features present. Experimentation with different parameters of the radially variable spatial filtering technique provides the best results among such techniques as it allows for increased kernel size at larger cometocentric distances from the nucleus.

To determine the spatial locations of features and thereby to make conclusions based on such measurements, such as the outflow velocities, one needs to be confident that the enhancement techniques used do not alter the actual spatial locations of the features either due to the removal of a background or image artifacts introduced to the coma feature as a result of the mathematical operations carries out on the original image during the enhancement process. Again, confirmation of results with different categories of techniques is advisable. Of course, low-level image artifacts present in images including those newly introduced or not corrected during the standard image reduction process such as the flat fielding, which may not have been successful in properly removing a gradient in the images, could get amplified during the enhancement process.

Application of multiple enhancing techniques in succession to a given image is extremely tempting and it has been attempted in the literature (e.g., Schwarz et al. 1989, Manzini et al. 2012). An example is the application of a technique suited for detecting radial features (such as division by a "background") followed by a technique that searches for "wiggles" or "inflections" in the radial features[3] to identify shell-like structures and such a technique could be as simple as dividing by an "ideal" radial profile (Schwarz et al. 1989). While this approach is reasonable, such shell-like structures should also be detectable by other techniques already discussed in this paper (e.g., radially variable spatial filtering, radial shift differencing, temporal derivative). Therefore, the golden rule applies again; confirm the features with an unrelated enhancement technique.

In recognition of the fact that relevant software to carry out some of the above image enhancement techniques is not available as open source, we have developed a web facility to mitigate this limitation. At the URL http://www.psi.edu/research/cometimen we provide a web interface for a user to upload an image of a cometary coma, to enhance it with certain image enhancement techniques discussed in this paper, and finally to download the enhanced image files (Samarasinha et al. 2013). The techniques accessible at the web facility are not available in the NOAO and ESO packages IRAF and MIDAS. At the facility, supplementary material such as a tutorial on how to use the techniques and the corresponding source codes are also available.

The online supplementary material in *Icarus* contains the simulated images used in the figures of this paper in FITS format.

# Acknowledgements

NHS acknowledges NASA Planetary Mission Data Analysis grant NNX09AM03G and NASA Planetary Atmospheres grant NNX12AG56G for supporting this work. We thank Ruvini Samarasinha for making schematic Figures 2, 3, A1, A2, and A3. We thank Dr. Matthew Knight and an anonymous reviewer for constructive comments that improved the quality of the paper. This is PSI Contribution Number 616.

---

[3] Manzini et al. (2012) call these "radial de-trending techniques".

# Appendix A

In this Appendix we show that (a) a steady-state spherically symmetric outflow at constant speed will result in an azimuthally symmetric $1/\rho$ column density distribution in the coma and (b) a steady-state jet of constant outflow speed originating from a non-rotating (or slow rotating) nucleus will result in an azimuthally *asymmetric* column density distribution; however, the corresponding radial column density distribution will have a $1/\rho$ fall-off.

(a) A steady-state spherically symmetric outflow at constant speed

Let $Q$ be the steady rate of emission of molecules (or particles) from the nucleus and $V$ the constant outflow speed. Then the spherically symmetric number density $n(R)$ at a distance $R$ from the nucleus (Fig A1) is given by

$$n(R) = \frac{Q}{4\pi R^2 V} \ . \tag{A1}$$

Then the column density $N(\rho)$ at a skyplane-projected distance $\rho$ from the nucleus is given by

$$N(\rho) = \int_{-\infty}^{\infty} \frac{Q}{4\pi R^2 V} dx \tag{A2}$$

where $x$ is the line-of-sight distance measured from the skyplane and $R$, $\rho$, $x$, and the angle $\eta$ between the skyplane and the point $x$ are related by

$$R = \rho \sec \eta \tag{A3}$$

and

$$x = \rho \tan \eta \ . \tag{A4}$$

Therefore,

$$dx = \rho \sec^2 \eta \, d\eta \tag{A5}$$

leading to

$$N(\rho) = \frac{Q}{4\pi V \rho} \int_{-\pi/2}^{\pi/2} d\eta \ . \tag{A6}$$

I.e., $N(\rho) = \dfrac{Q}{4V} \dfrac{1}{\rho}$ \hfill (A7)

and

$$N(\rho) \propto \frac{1}{\rho} \ .$$

(b) A steady-state outflow in the form of a jet of constant outflow speed

First consider the case where the radial column density profile $N(\rho)$ is measured along the skyplane-projected nucleus–jet axis line (Fig A2).

Let the number density $n(R)$ at a distance $R$ from the nucleus and at an angle $\kappa$ away from the jet axis be given by

$$n(R) = \frac{Q}{V} \frac{f(\kappa)}{R^2} \tag{A8}$$

where $Q$ is the steady rate of emission of molecules (or particles) from the nucleus, $V$ is the constant outflow velocity, $f(\kappa)$ is the term describing the number density dependence as a function of angle $\kappa$ as one moves away from the jet axis. Let the jet axis be at an angle $\lambda$ away from the skyplane.

Then the column density $N(\rho)$ at a skyplane-projected distance $\rho$ from the nucleus along the skyplane-projected nucleus–jet axis line is given by

$$N(\rho) = \int_{-\infty}^{\infty} \frac{Q}{V} \frac{f(\kappa)}{R^2} dx \tag{A9}$$

where $x$ is the line-of-sight distance measured from the skyplane and $R$, $\rho$, $x$ and the angles $\kappa$, $\lambda$ and $\eta$ are related by

$$R = \rho \sec \eta, \tag{A10}$$
$$x = \rho \tan \eta \tag{A11}$$

and

$$\eta = \lambda - \kappa. \tag{A12}$$

Therefore,

$$dx = \rho \sec^2 \eta \, d\eta \tag{A13}$$

and

$$d\eta = -d\kappa \tag{A14}$$

leading to

$$N(\rho) = \frac{Q}{V\rho} \int_{\lambda - \pi/2}^{\lambda + \pi/2} f(\kappa) d\kappa. \tag{A15}$$

Note that the integral is independent of $\rho$.

Therefore, $N(\rho) \propto \dfrac{1}{\rho}$.

Now consider the case where the skyplane projected jet axis is offset by an angle $\gamma$ to the line along which the radial column density profile is measured (Fig A3).

Then the number density $n(R)$ at a distance $R$ from the nucleus and at an angle $\chi$ away from the jet axis (analogous to equation A8) is given by

$$n(R) = \frac{Q}{V} \frac{f(\chi)}{R^2} \tag{A16}$$

resulting in a column density $N(\rho)$ at a skyplane projected distance $\rho$ from the nucleus of

$$N(\rho) = \int_{-\infty}^{\infty} \frac{Q}{V} \frac{f(\chi)}{R^2} dx. \tag{A17}$$

Here $x$ is the line-of-sight distance measured from the skyplane and $R$, $\rho$, $x$, and the angle $\eta$ are related by

$$R = \rho \sec \eta \tag{A18}$$

and
$$x = \rho \tan \eta \ . \tag{A19}$$
Therefore,
$$N(\rho) = \frac{Q}{V\rho} \int_{-\pi/2}^{\pi/2} f(\chi) d\eta \ . \tag{A20}$$
However, by considering the dot product,
$$\cos \chi = \frac{\rho^2 + x\,l}{s\,R} \ . \tag{A21}$$
Since $l$, $s$, $\rho$, and the angles $\lambda$ and $\gamma$ are related by
$$l = \rho \tan \lambda \tag{A22}$$
and
$$s^2 = \rho^2 + (\rho \tan \gamma)^2 + (\rho \tan \lambda)^2 \ , \tag{A23}$$
equation (A21) can be rewritten as
$$\cos \chi = \frac{1 + \tan \eta \tan \lambda}{\sec \eta (1 + \tan^2 \gamma + \tan^2 \lambda)^{1/2}} = \frac{\cos(\lambda - \eta)}{(1 + \cos^2 \lambda \tan^2 \gamma)^{1/2}} \ . \tag{A24}$$
As angles $\lambda$ and $\gamma$ are fixed, $\chi$ is a function of $\eta$ and the integral in the equation (A20) is a constant. I.e., $N(\rho) \propto \frac{1}{\rho}$. However, note that the column density is not azimuthally symmetric in the skyplane as it depends on the angle $\gamma$.

# Appendix B

In this Appendix, error analyses for the enhanced images corresponding to division by an $1/\rho$ profile and subtraction of an $1/\rho$ profile, respectively are presented. Based on this, we conclude that for everywhere in the enhanced image, the division by an $1/\rho$ profile yields better diagnostic results for identification of coma features. At every location, it has smaller relative noise (i.e., noise-to-signal ratio) than those corresponding to the subtraction of an $1/\rho$ profile.

Let $\delta I_{in}(x,y)$ be the noise associated with a given pixel $(x,y)$ having a pixel value $I_{in}(x,y)$ in the image to be enhanced.

## Division by an $1/\rho$ profile:

Division by an $1/\rho$ profile is expressed by
$$I_{out,div}(x,y) = I_{in}(x,y) \rho(x,y) \tag{B1}$$
where $I_{out,div}(x,y)$ is the pixel value of the pixel $(x,y)$ of the enhanced image.

Then the noise of a given pixel in the enhanced image, $\delta I_{out,div}(x,y)$, can be expressed by

$$\delta I_{out,div} = \frac{\partial I_{out,div}}{\partial I_{in}} \delta I_{in} = \rho \, \delta I_{in} \tag{B2}$$

since $\rho(x,y)$ is not a variable here.

Therefore the relative noise is given by

$$\frac{\delta I_{out,div}}{I_{out,div}} = \frac{\delta I_{in}}{I_{in}} \; . \tag{B3}$$

## Subtraction of an $1/\rho$ profile:

Subtraction of an $1/\rho$ profile is expressed by

$$I_{out,sub}(x,y) = I_{in}(x,y) - \frac{k}{\rho(x,y)} \tag{B4}$$

where $I_{out,sub}(x,y)$ is the pixel value of the pixel $(x,y)$ of the enhanced image and $k$ is the normalization constant.

Then the noise of a given pixel $(x,y)$ in the enhanced image, $\delta I_{out,sub}(x,y)$, can be expressed by

$$\delta I_{out,sub} = \frac{\partial I_{out,sub}}{\partial I_{in}} \delta I_{in} = \delta I_{in} \tag{B5}$$

since $\rho(x,y)$ is not a variable here.

Therefore the relative noise is given by

$$\frac{\delta I_{out,sub}}{I_{out,sub}} = \frac{\delta I_{in}}{\left(I_{in} - \frac{k}{\rho}\right)} \; . \tag{B6}$$

Therefore, based on (B3) and (B6), for any pixel $(x,y)$ we have

$$\frac{\delta I_{out,div}}{I_{out,div}} < \frac{\delta I_{out,sub}}{I_{out,sub}} \; . \tag{B7}$$

For all other enhancement techniques in this paper where both the division by a background profile and the subtraction of a background profile are feasible, using analogous approaches to that in this Appendix, one can show that division will result in smaller relative noise than the subtraction.

Asteroids, Comets, Meteors 1991. Published by Lunar and Planetary Institute, Houston, Texas, USA.

Lederer, S.M., Campins, H., and Osip, D.J., 2009. Chemical and Physical Properties of Gas Jets in Comets. II. Modeling OH, CN and $C_2$ Jets in Comet C/1995 O1 (Hale Bopp) One Month after Perihelion. Icarus, 199, 484-504.

Lorenz, H., Richter, G.M., Capaccili, M., and Longo, G., 1993. Adaptive Filtering in Astronomical Image Processing I. Basic Considerations and Examples. Astron. Astrophys., 277, 321-330.

Manzini, F., Behrend, R., Comolli, L., Oldani, V., Cosmovici, C.B., Crippa, R., Guaita, C., Schwarz, G., and Coloma, J., 2012. Comet Machholz (C/2004 Q2): Morphological Structures in the Inner Coma and Rotation Parameters. Astrophys. Space Sci., 337, 531-542.

McCarthy, Jr., D.M., Stolovy, S.R., Campins, H., Larson, S., Samarasinha, N.H., and Kern S.D., 2007. Comet Hale-Bopp in Outburst: Imaging the Dynamics of Icy Particles with HST/NICMOS. Icarus, 189, 184-195.

Mueller, B.E.A., Samarasinha, N.H., Farnham, T.L., and A'Hearn, M.F. 2013. Analysis of the Sunward Continuum Features of Comet 103P/Hartley 2 from Ground-based Images. Icarus, 222, 799-807.

Murtagh, F., Zeilinger, W.W., Stark, J.-L., and Boehnhardt, H. 1993. Detection of Faint Extended Structures by Multiresolution Wavelet Transformation. The Messenger, 73, 37-39.

O'Gorman, L., Sammon, M.J., and Seul, M., 2008. Practical Algorithms for Image Analysis. Cambridge University Press, Cambridge, UK.

Polesel, A., Ramponi, G., and Mathews, V.J. 2000. Image Enhancement Via Adaptive Unsharp Masking. IEEE Transactions on Image Processing, 9(3), 505-510.

Samarasinha, N.H., and Larson, S.M., 2011. Comparison of Image Enhancement Techniques for Cometary Comae. EPSC-DPS Joint Meeting 2011, 1400.

Samarasinha, N.H., Martin, M.P., and Larson, S.M., 2013. Cometary Coma Image Enhancement Facility. http://www.psi.edu/research/cometimen.

Schleicher, D.G. and Farnham, T.L., 2004. Photometry and Imaging of the Coma with Narrowband Filters. In Comets II (Eds: M.C. Festou, U. Keller, H.A. Weaver), University of Arizona Press, Tucson, Arizona, USA.
30

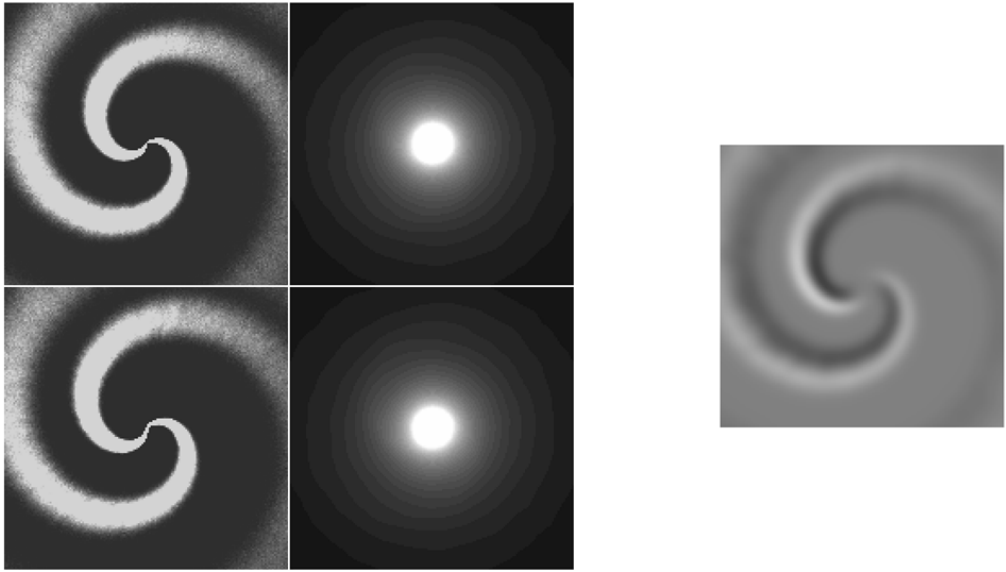

Figure 1. Above simulations demonstrate the temporal derivative technique. On the leftmost column are two jet features emanating from the nucleus, located at the center of each panel, with the bottom panel showing the jet structure 0.05 of a rotation period later. The second column shows the respective coma structure after addition of a background coma and convolution of the simulated images by a Gaussian to mimic the astronomical seeing. The panel on the extreme right is the resultant image when the later image is divided by the earlier image. The bright (white) jet structure corresponds to the spatial locations of the jets in the later image whereas the darker structure corresponds to the respective spatial locations in the earlier image. Note that the widths of the jets are not preserved in the temporal derivative; the widths of the structures in the temporal derivative depend on the actual widths of the features as well as on their relative displacement between the early and later images.

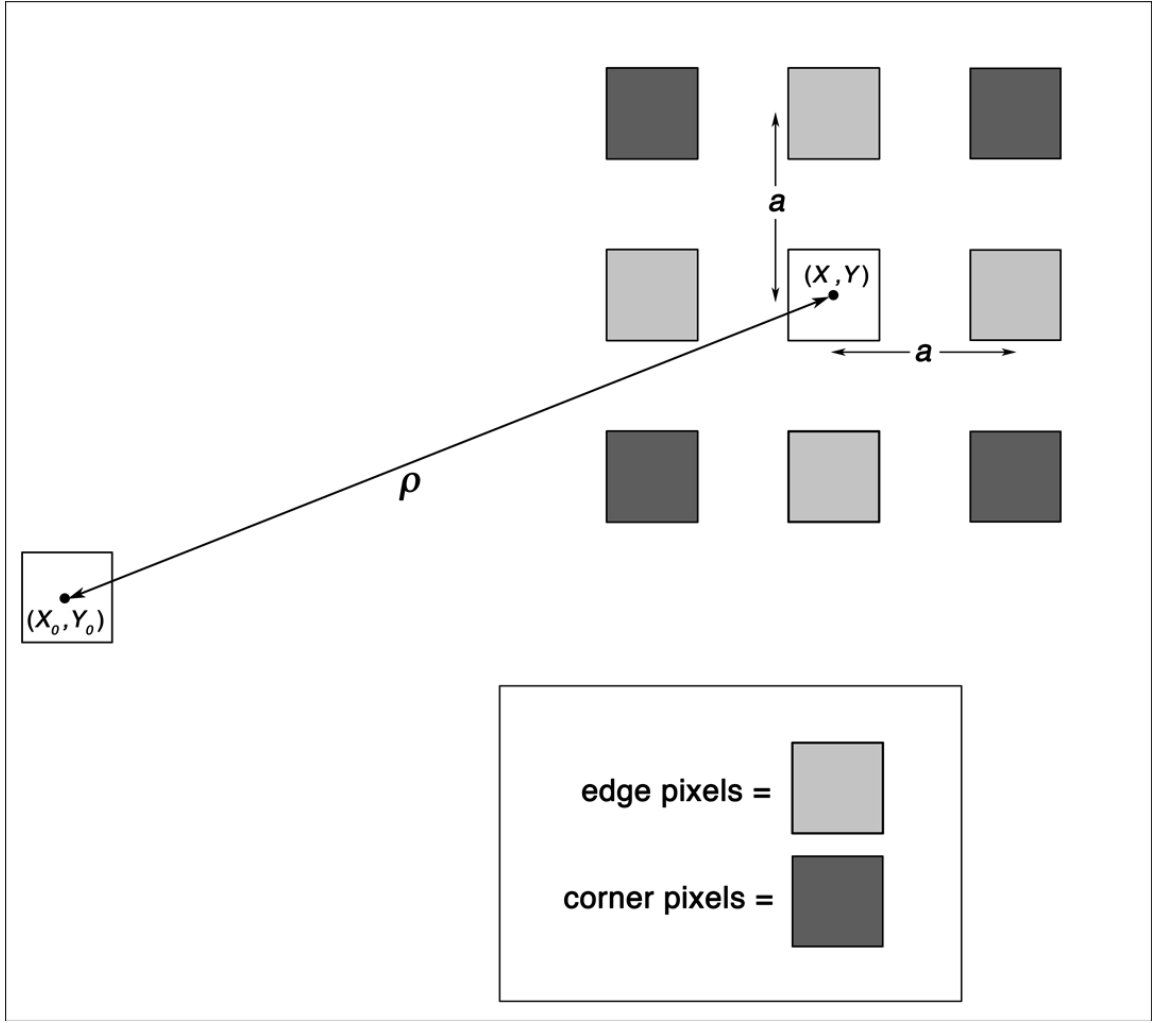

Figure 2. A schematic diagram showing the 3×3 orthogonal pixel matrix centered at the target pixel at (*x,y*). The target pixel is located at a distance $\rho$ from the reference pixel at ($x_0, y_0$). The separation of the edge pixels from the target pixel increases as the distance $\rho$ increases. The radially variable spatial filtering technique will determine the new pixel value of the target pixel and is based on the individual pixel values of the 3×3-pixel matrix.

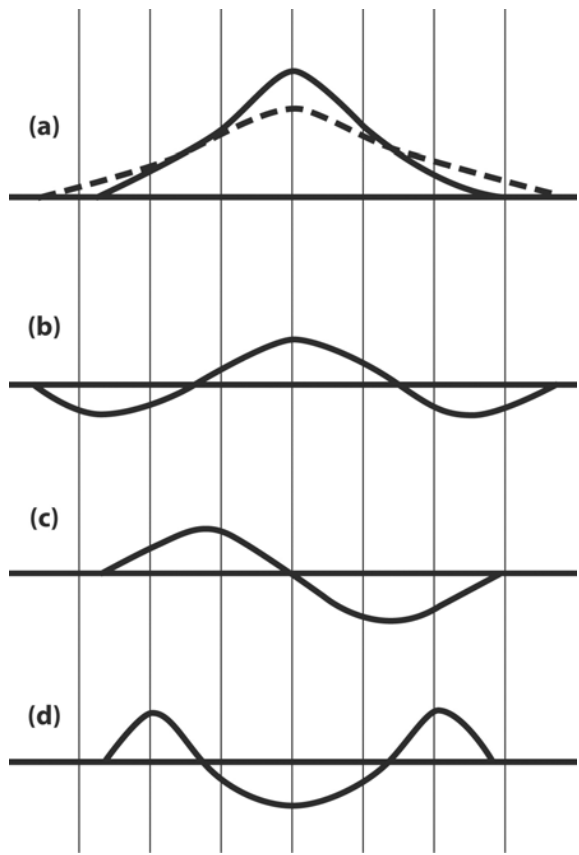

Figure 3. A cartoon showing the graphical representations for certain operations related to enhancement techniques. (a) cross section of a jet is shown as a solid line where the x-axis denotes the spatial dimension and the y-axis represents the brightness. The dashed line represents the cross section of the same jet after being convolved with a Gaussian kernel. For clarity, only the one-dimensional case is shown. (b) after application of the unsharp masking or a variation of it (e.g., division of the original image by the Gaussian-convolved image), the peak brightness of the jet is present in the same location. However, two "holes" are created on either side of the jet. (c) first derivative of the jet shown in the top panel. (d) second derivative of the jet shown in the top panel. The second derivative techniques such as Laplace filtering create two jets at the edges in addition to a "hole" in the middle. Care must be exercised in interpretation of images enhanced with such techniques.

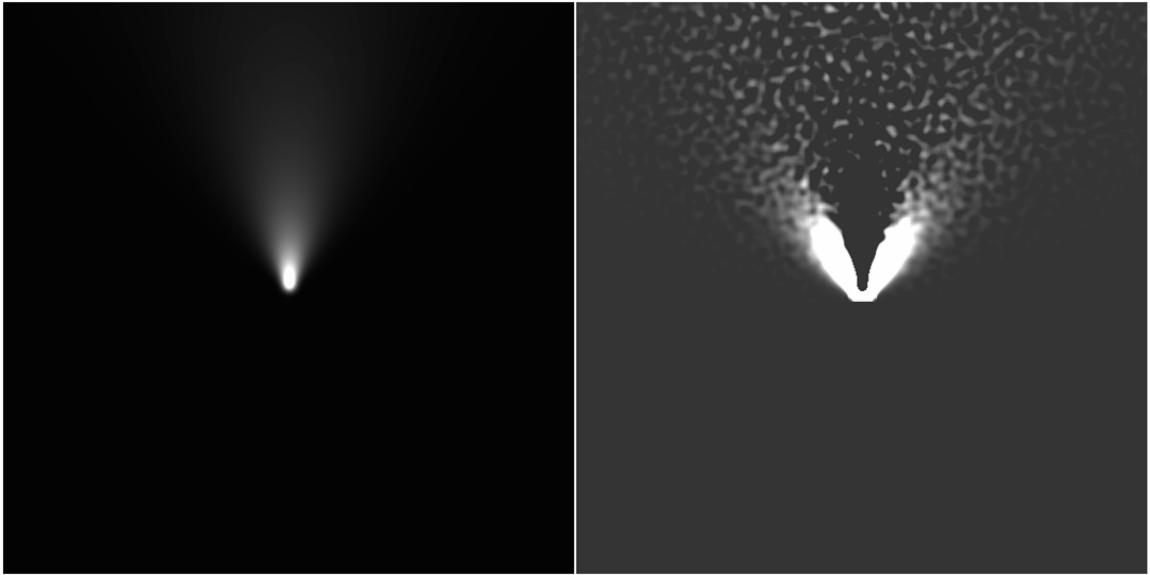

Figure 4. An unenhanced image depicting a single simulated jet (left) and the resultant image after applying the Laplace filter (right). As demonstrated in Fig 3, two "spurious" jets are created at either edge of the actual jet. White denotes brighter regions. As seen in the examples to follow, when multiple jet features are present, the Laplace filtered images may lead one to incorrectly interpret the locations and the number of jet features.

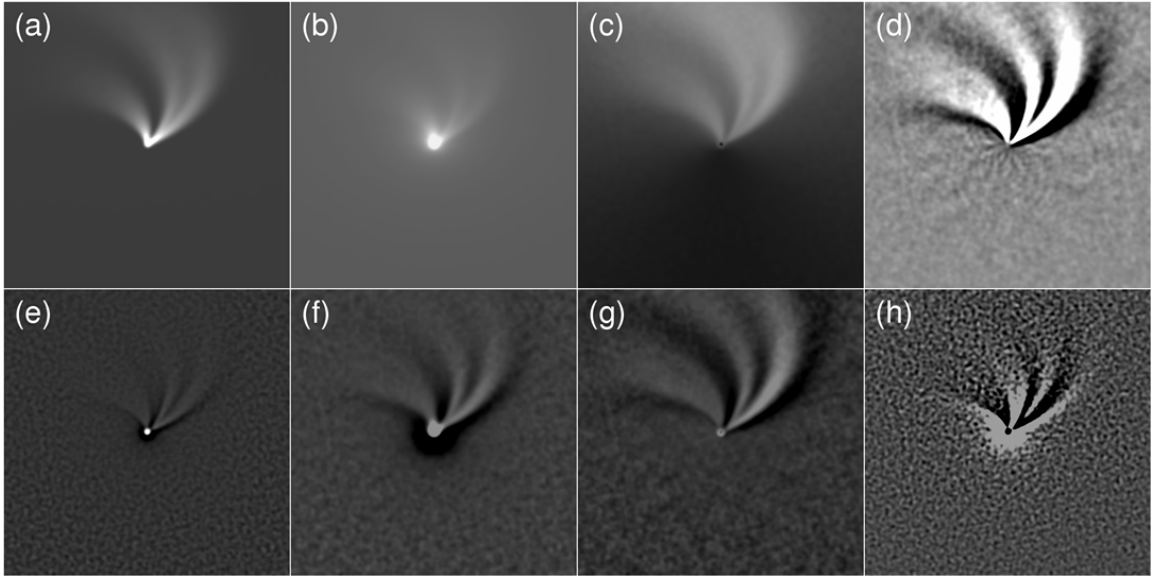

Figure 5. The panels in this figure compare the effects of various enhancement techniques. In contrast to real observations, the three simulated jets used in this figure provide a priori knowledge of the jets present and their locations and therefore a realistic assessment of the enhanced images can be made. From left to right in the top row are: (a) an unenhanced but highly stretched image depicting the three jets mentioned above convolved with a Gaussian to mimic the astronomical seeing, (b) again an unenhanced but highly stretched image where the same three jets are added to a broad asymmetric background corresponding to hemispherical emission (the sun is towards the top) convolved with a Gaussian to mimic the astronomical seeing, (c) image shown in the previous panel after division by an $1/\rho$ profile, and (d) application of the rotational shift differencing technique with 10° each clockwise and counterclockwise rotations for the image shown in panel (b). From left to right in the bottom row are: (e) division of the image shown in panel (b) by a Gaussian convolved image — a technique that can be considered as a variant of the classical unsharp masking; standard deviation of the Gaussian is 3 pixels, (f) the same enhancement as the previous panel but convolved with a Gaussian of standard deviation 15 pixels to demonstrate that as the kernel size of the smoothing function increases, large-scale spatial structures further out are enhanced whereas the smaller spatial-scale features close to the nucleus lose resolution, (g) after application of a radially variable spatial filter to image in panel (b), and (h) after application of a Laplace filter to the image in panel (b). Note that identification of features that mimic jets in the Laplace filtered image could lead to possible misinterpretations.

In this figure and for all subsequent figures, the nucleus is at the center of each panel and the contrast is chosen to highlight as many features as possible. White denotes brighter regions/features in the coma.

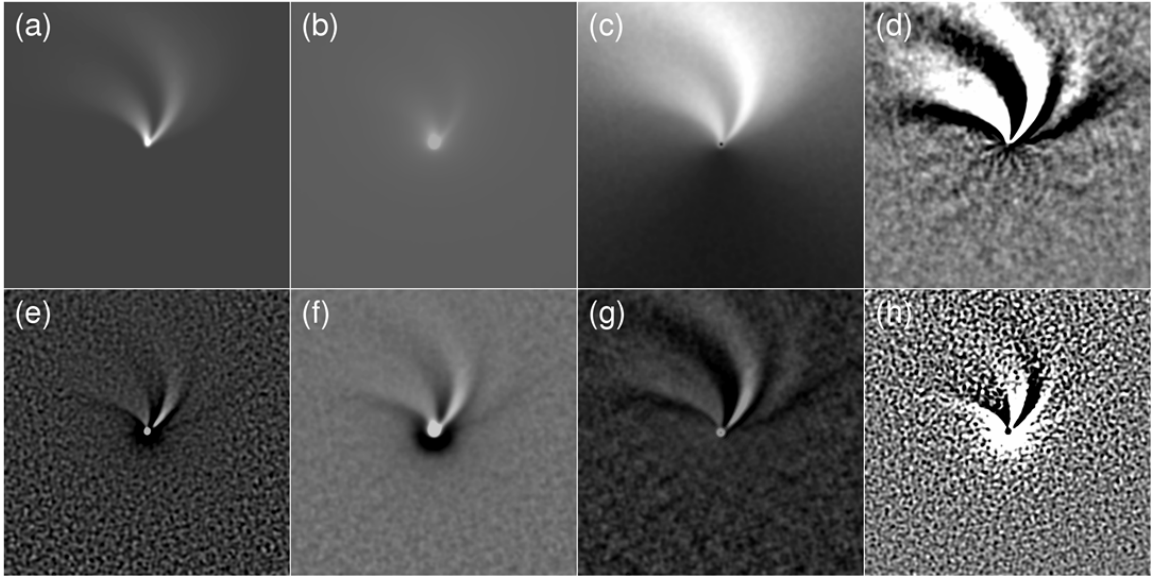

Figure 6. The same as in Figure 5 except that the top rightmost jet is now significantly weaker. From left to right and top to bottom are: (a) an unenhanced but highly stretched image showing the three jets convolved with a Gaussian to mimic the astronomical seeing; note that the rightmost jet is now barely discernible from the middle jet, (b) after addition of the asymmetric background and convolution with a Gaussian to mimic the astronomical seeing, (c) after division of the image in panel (b) by an $1/\rho$ profile, (d) after application of the rotational shift differencing, (e) and (f) after division by a Gaussian convolved image with a standard deviation of 3 pixels and 15 pixels respectively, (g) after application of a radially variable spatial filter, and (h) after application of a Laplace filter. Note that identification of features that mimic jets in the Laplace filtered image could still lead to possible misinterpretations; however, the radially variable spatial filter as well as the rotational shift differencing is capable of detecting the weak rightmost jet.

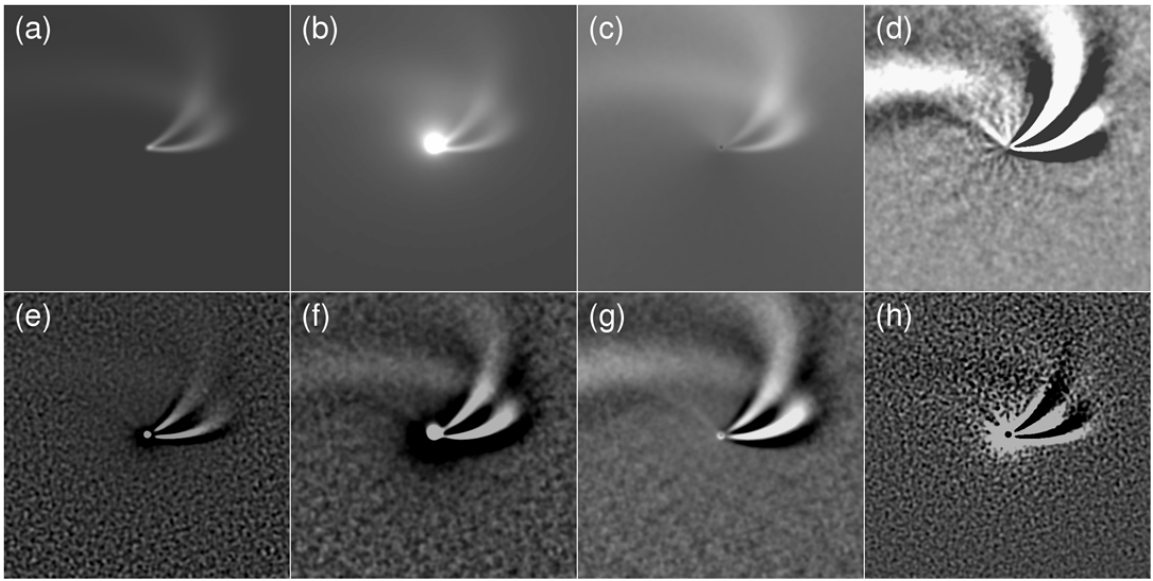

Figure 7. Similar to Figures 5 and 6 except that now there are two overlapping jets. From left to right in the top row are: (a) an unenhanced but highly stretched image depicting two overlapping jets convolved with a Gaussian, (b) again an unenhanced but stretched image where the two overlapping jets are added to the same asymmetric background used in Figures 5 and 6 and convolved with a Gaussian, (c) after division of (b) by an $1/\rho$ profile, (d) after application of the rotational shift differencing technique. From left to right in the bottom row are: (e) and (f) after division by a Gaussian convolved image with standard deviation of 3 pixels and 15 pixels respectively, (g) after application of a radially variable spatial filter, and (h) after application of a Laplace filter.

Note that the division by an $1/\rho$ profile and the radially variable spatial filter provide the best results in detecting the overlapping jets. On the other hand, rotational shift differencing gives a faulty impression. The broad upward feature just above the nucleus in panel (d), where a hint of which is also seen in some other panels, is due to the asymmetric background. Unlike in Figures 5 and 6, there is no strong upward jet(s) to mask the contribution from the asymmetric background and therefore it becomes more apparent in this case.

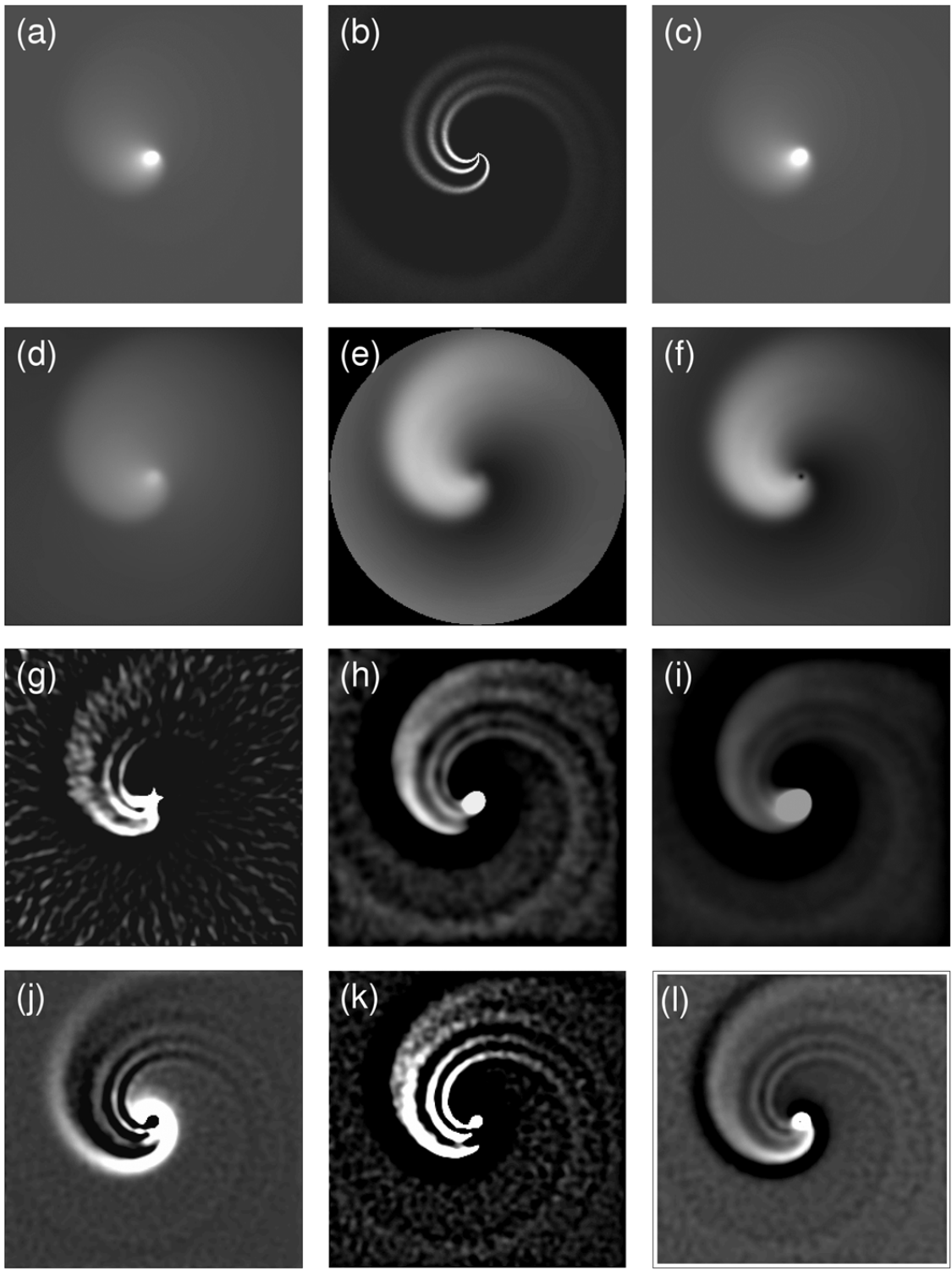

Figure 8. The comparative abilities of different enhancement techniques to recover overlapping coma features with different brightness levels. From left to right in the top row are: (a) a simulation showing a broad jet originating from a large source region on the nucleus, (b) a simulation of three narrow features originating from the same large source region (the broad jet in the previous panel is approximately two orders of magnitude brighter than these narrow features; the brightness stretch used in this panel is different to that used in the previous panel), and (c) after co-adding the two simulated images in panels (a) and (b) and convolving the resultant image by a Gaussian filter to mimic astronomical seeing (this panel is equivalent to a broad jet with small scale substructure within it). The next panels demonstrate the results after applying various image enhancement techniques to this image. From left to right in the second row are: (d) display of the simulated image in panel (c) using a logarithmic scale (note: the azimuthal profiles are calculated only within the full complete circles when pixels are present for all azimuths for a given radius and therefore the pixels near the corners of the panels are not enhanced), (e) after dividing the simulated image in panel (c) by its azimuthally averaged image, and (f) after division of the simulated image by $1/\rho$ where $\rho$ is the skyplane-projected cometocentric distance. The panels in the third row from left to right are: (g) after rotational shift differencing the simulated image in panel (c) by one degree, (h) after convolving the simulated image in panel (c) by a Gaussian of 7-pixel standard deviation and then dividing the image in panel (c) by the convolved image — a variant of the classical unsharp masking, and (i) a similar enhancement to the that in the previous panel but the convolution carried out with a 15-pixel Gaussian filter. The panels in the bottom row from left to right are: (j) after application of the Laplace filter, (k) an inverted display after application of the Laplace filter, and (l) after application of the radially variable spatial filter.

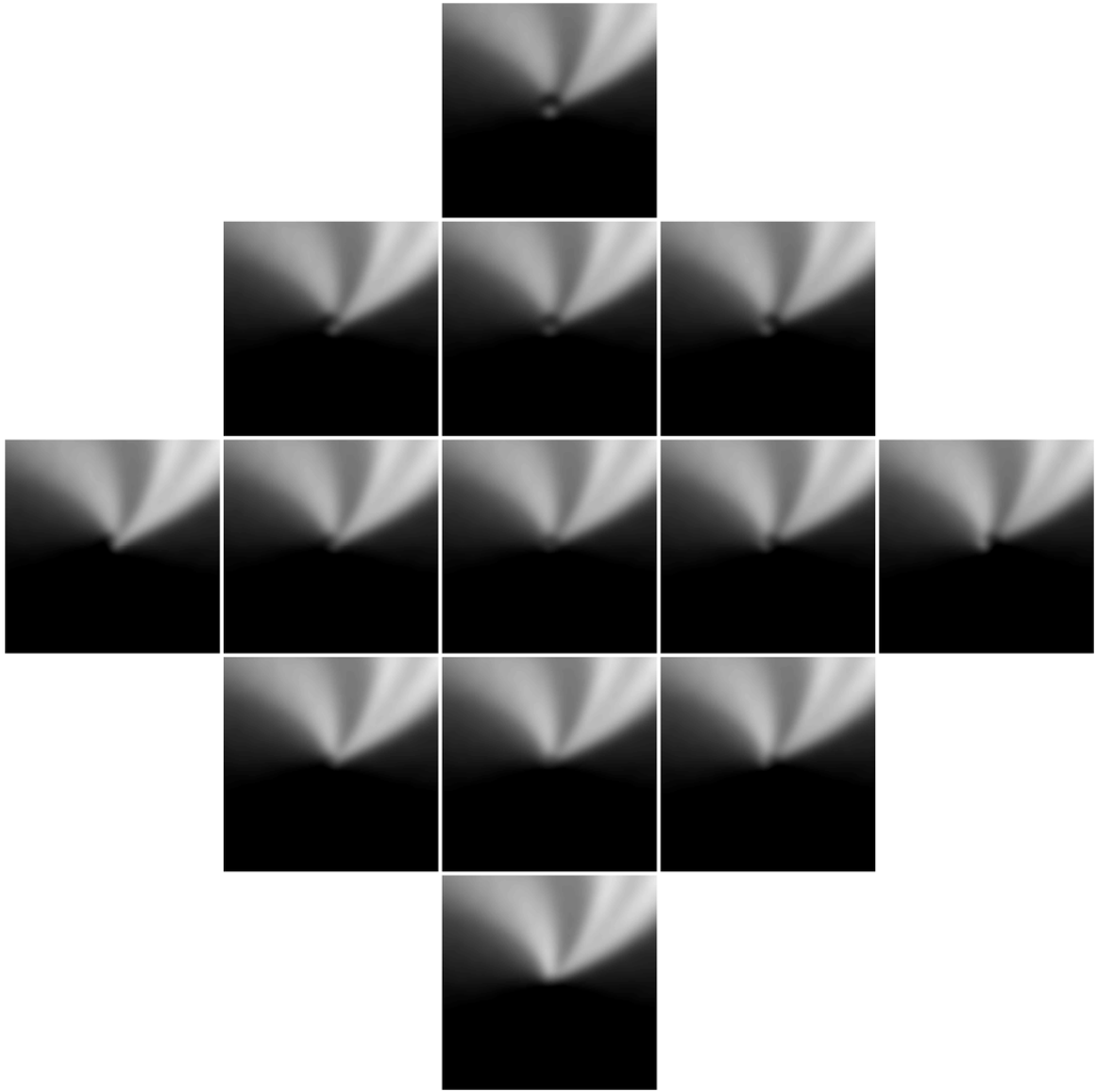

Figure 9. Effects on the morphology near the nucleus in enhanced images when the adopted image center is offset from the optocenter are shown above. The inner portion of the simulated image shown in figure 5 is used here. All the panels are enhanced using the division by azimuthal average technique. The center panel corresponds to the case when the optocenter is chosen as the nucleus. The chosen nucleus is offset by one pixel from the optocenter for panels immediately above, below, left, or right from the center panel. The offset (from the optocenter) for the nucleus is one pixel between any two adjacent vertical or horizontal panels and is in the direction from the center panel. Convolving the original simulated image with a Gaussian prior to the image enhancement mimics the astronomical seeing. Notice that the coma morphologies are different from each other near the chosen nucleus for different panels and this clearly demonstrates that the near-nucleus morphology is sensitive to the chosen location for the nucleus.

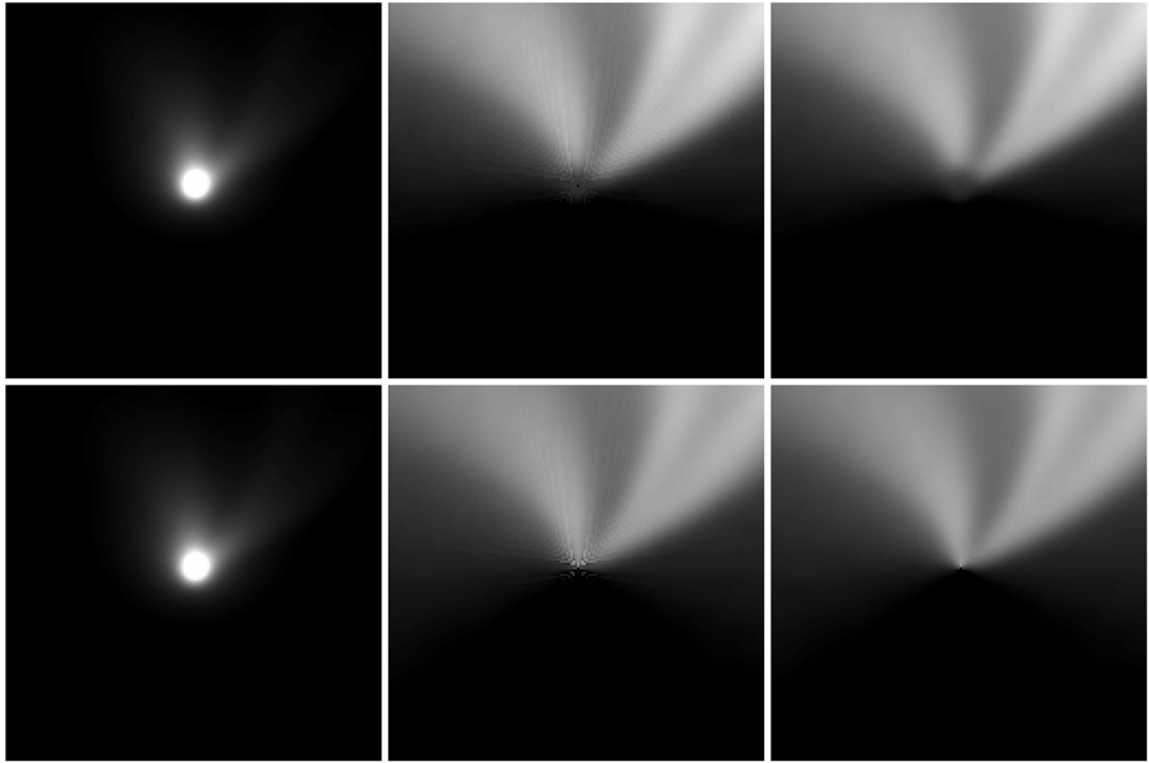

Figure 10. The enhancement techniques that entail determination of the location of the nucleus and the subsequent conversion of the rectangular coordinates into polar coordinates require sub-pixel sampling and appropriate weighting to avoid image artifacts near the nucleus. Shown on the left is the inner portion of the simulated image used in Figure 5. The second column shows the application of division by azimuthal average (top) and azimuthal renormalization (bottom) without any sub-pixel sampling or weighting. The last column shows the application of division by azimuthal average (top) and azimuthal renormalization (bottom) but with sub-pixel sampling and corresponding weighting. Each pixel is sampled into 10×10 sub-pixels. The "cross-like" artifact seen in the panels of the second column, for both techniques, is due to poor sampling of pixels near the nucleus.

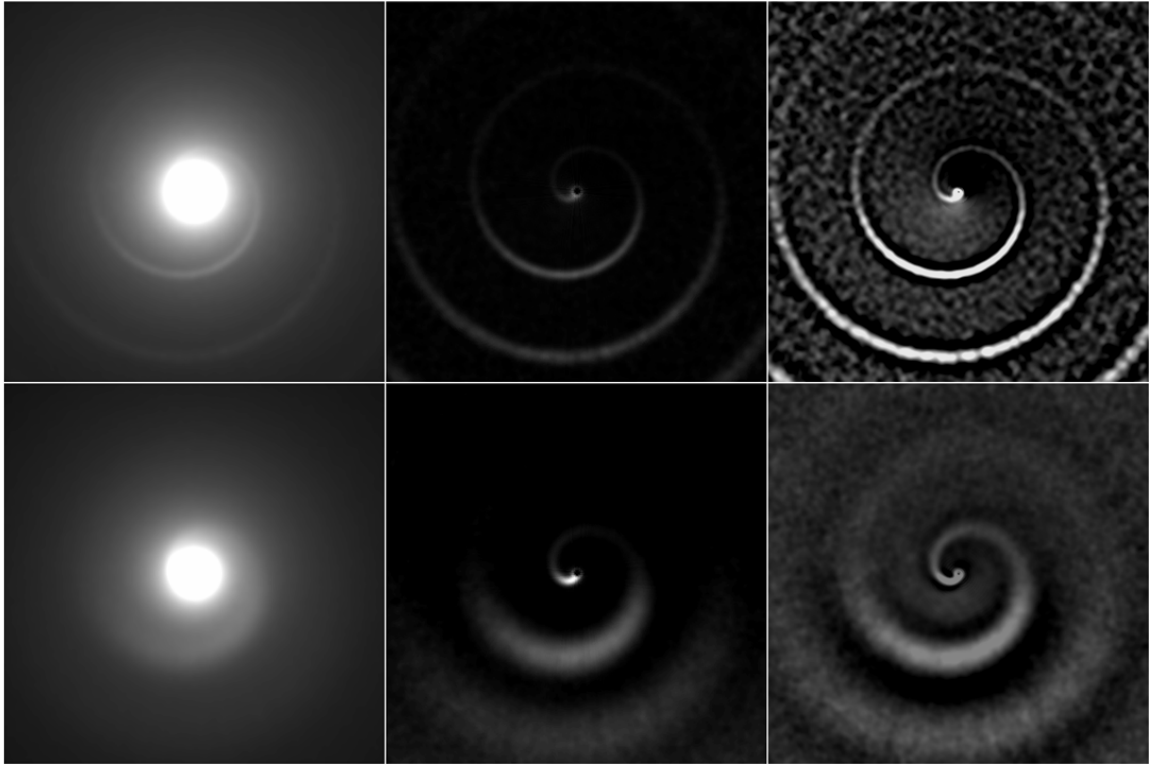

Figure 11. Top: From left to right are (a) an unenhanced narrow Archimedean spiral of variable brightness superimposed on a uniform background corresponding to a steady state spherically symmetric emission and then convolved with a Gaussian to mimic astronomical seeing, (b) after division by a $1/\rho$ profile, and (c) after application of a radially variable spatial filter. Bottom: Same as top except for a wide jet. For each panel, the nucleus is at the center and the contrast is arbitrarily chosen to show the features.

Sets of images corresponding to different times of observations (image sequences are not shown) are used to calculate the expansion velocity.

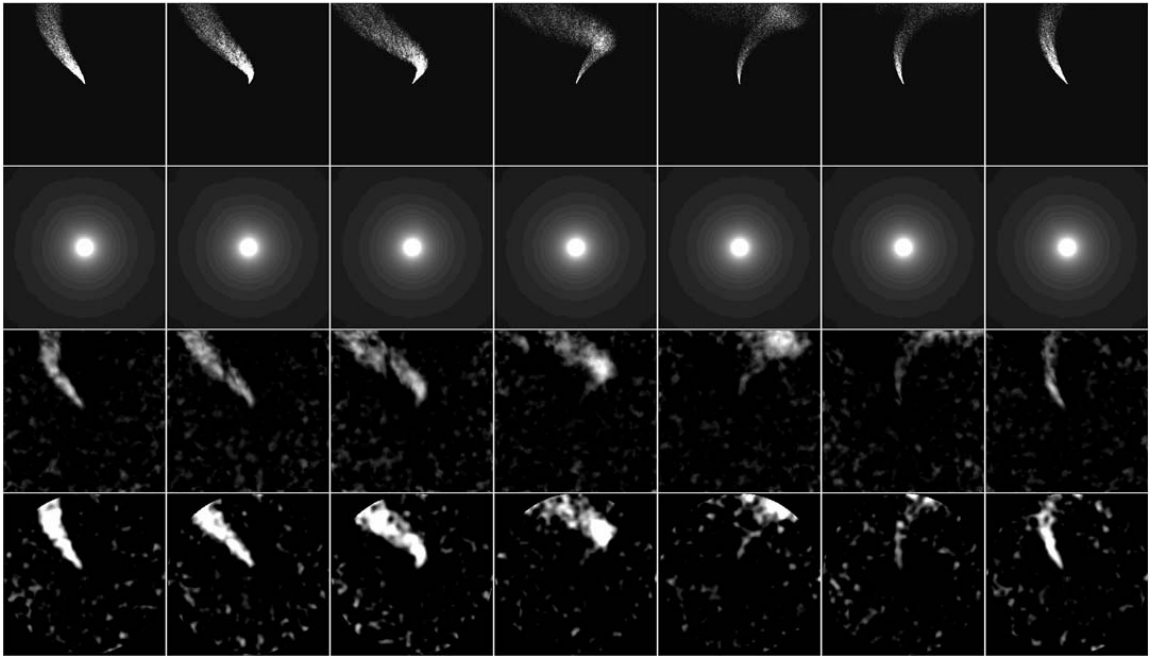

Figure 12. Shown left to right is a temporal sequence of simulated images covering a full rotation cycle. Images are simulated for rotational phases of 0.00, 0.15, 0.26, 0.44, 0.61, 0.71, and 0.87 providing an approximately equi-spaced sequence in the temporal domain, mimicking a likely sequence of actual observations. A low-contrast single jet feature and a strong background coma are simulated. The jet feature is only of the order a few percent of the strong background and is only comparable to or a little stronger than the signal-to-noise. The top row shows the simulated jet feature as a function of rotational phase while the second row shows the resultant simulated images when the jet feature is combined with the strong background coma and convolved with a Gaussian to mimic the astronomical seeing. These seven images are used to determine the median-mask image and the third row shows the results when this median mask is used to divide the images shown in the second row. The bottom row shows the corresponding images when the images in the second row are enhanced using the division by the azimuthally averaged profiles. The azimuthal profiles are calculated only within the full complete circles (i.e., when pixels are present for all azimuths for a given radius) and therefore the pixels near the corners of the panels are not enhanced. It is apparent that the division by the median mask generated from the sequence of images over a rotational cycle yields consistent results to what one derives from division by azimuthal average as well as to the original locations of jet features. However, it should be noted that the low-contrast nature of the simulated jets and the fact they are just above the signal-to-noise level in the second row of images makes the determination of the exact jet morphology uncertain in some cases (e.g., columns 4 and 5).

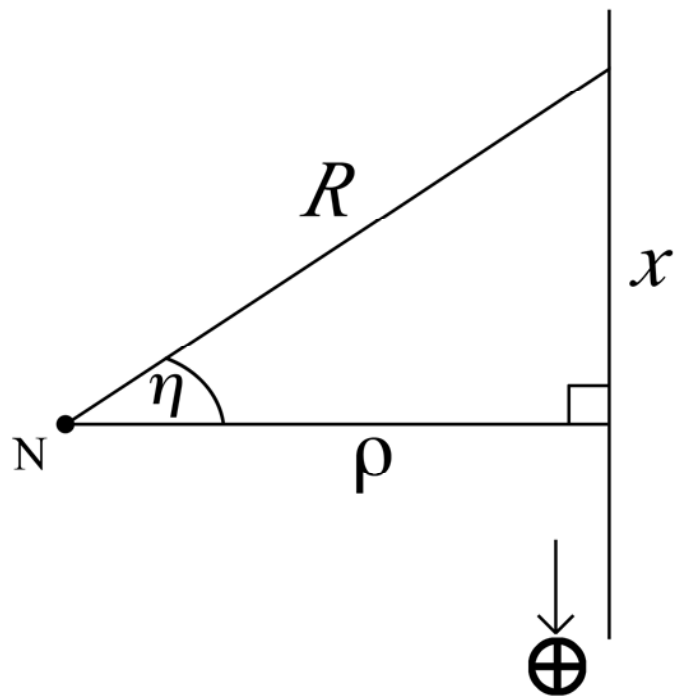

Figure A1. A schematic diagram corresponding to the spherically symmetric outflow from the nucleus showing the nucleus (N), the vertical line representing the line-of-sight, and the skyplane-projected distance $\rho$ from the nucleus to the line-of-sight. The Earth symbol denotes the direction to the Earth (or the observer).

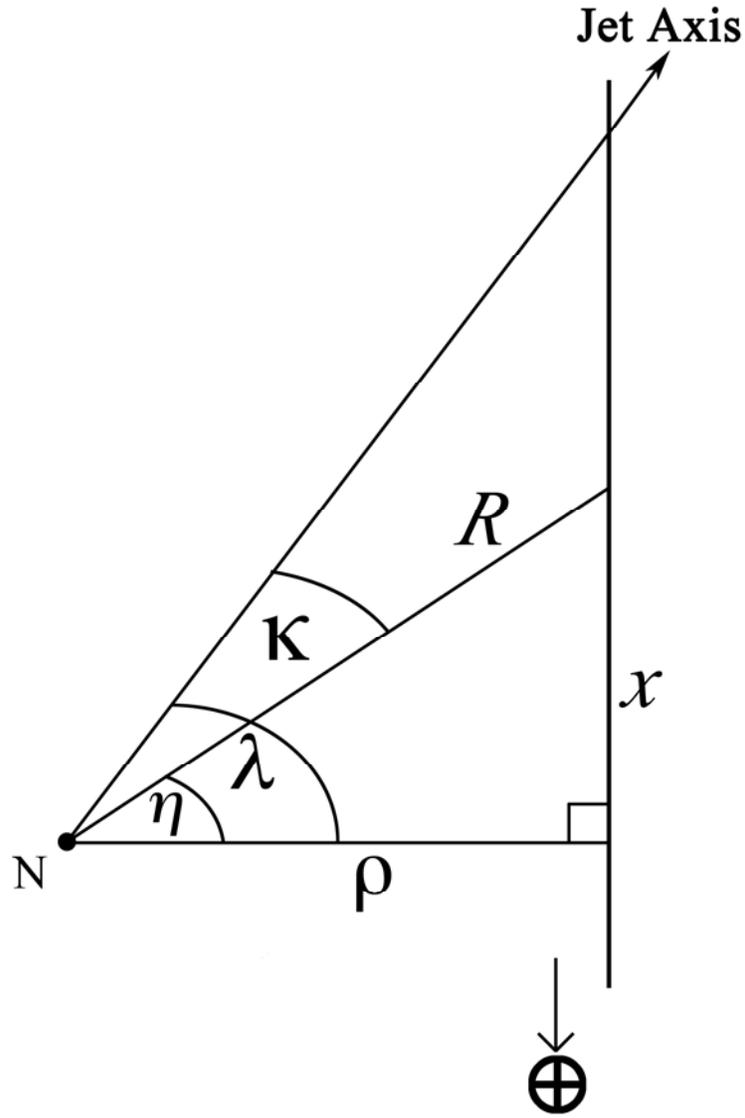

Figure A2. A schematic diagram corresponding to the case when the outflow is in the form of a jet and the column density is measured along the skyplane-projected nucleus-jet axis line. It depicts the nucleus (N), the vertical line representing the line-of-sight, and the skyplane-projected distance $\rho$ from the nucleus to the line-of-sight. The axis of the jet is at an angle $\lambda$ away from the skyplane. The Earth symbol denotes the direction to the Earth (or the observer).

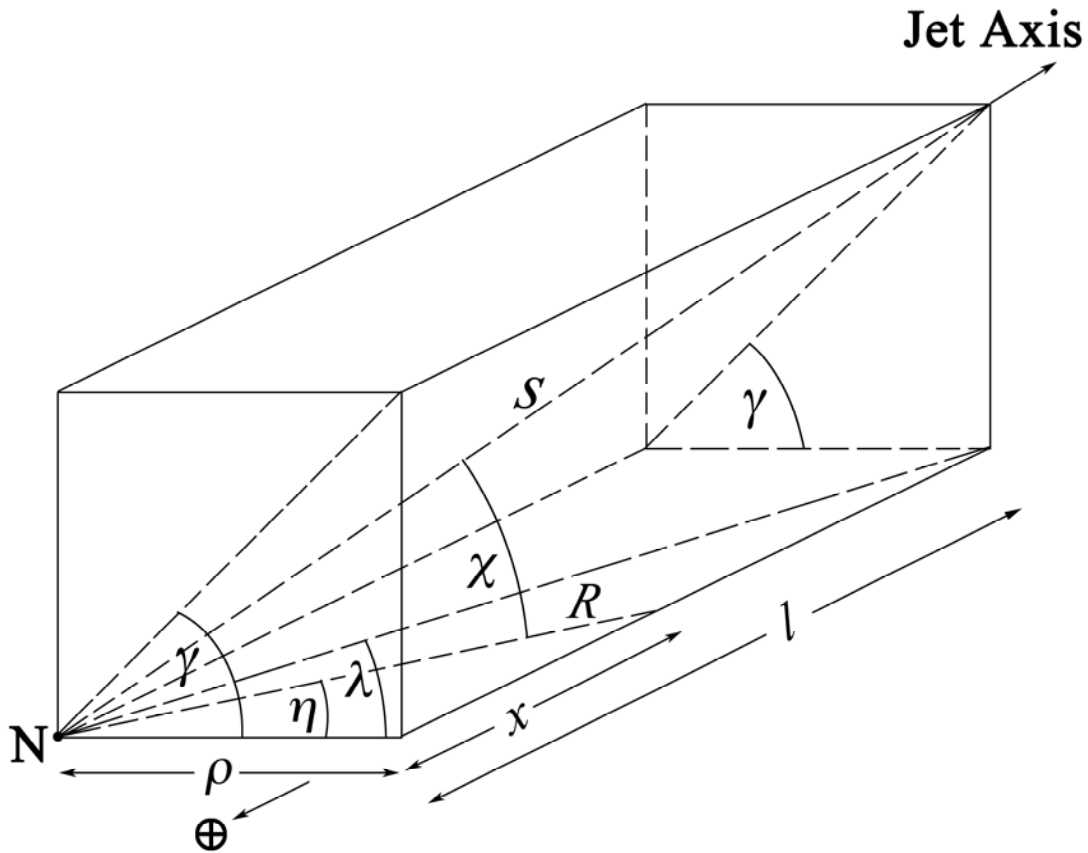

Figure A3. A schematic diagram corresponding to the case when the outflow is in the form of a jet and the column density is measured along a direction other than the skyplane-projected nucleus-jet axis line. It depicts the nucleus (N), the line representing the line-of-sight (along which the distance $x$ is measured), and the skyplane-projected distance $\rho$ from the nucleus to the line-of-sight. The axis of the jet when projected to the skyplane, makes an angle $\gamma$ with the direction that the column density is measured. The Earth symbol denotes the direction to the Earth (or the observer). When $\gamma = 0$, this case degenerates to that is shown in Fig A2.